%% file: article.tex
\documentclass[fleqn,usenatbib]{mnras}

\input{preamble}

\newcommand{\acceptancedate}{\DTMdate{2024-12-09}}
\newcommand{\receiveddate}{\DTMdate{2024-11-25}}
\newcommand{\originaldate}{\DTMdate{2024-02-21}}
\newcommand{\advpubdate}{\DTMdate{2025-01-07}}

\title[Precision matrix shrinkage estimators]{%
    A comparison of shrinkage estimators of the cosmological precision matrix%
}
\author[M.~J.~Looijmans, M.~(S.)~Wang and F.~Beutler]{%
    \parbox[b]{0.99\linewidth}{%
        \addauthor{Marnix~J.~Looijmans}%
            {Edinburgh}{}[0009-0003-0264-1274][mj.looijmans@tum.de]%
        \addauthor*{Mike~(Shengbo)~Wang}%
            {Edinburgh}{}[0000-0002-2652-4043]
        { and }%
        \addauthor*{Florian~Beutler}%
            {Edinburgh}{}[0000-0003-0467-5438]
    }
    \\%
    \parbox[b]{0.99\linewidth}{%
        \addaffil{Edinburgh}{}{Institute for Astronomy, University of Edinburgh, Royal Observatory Edinburgh, Blackford Hill, Edinburgh EH9~3HJ, UK}
    }%
}
\date{Accepted \acceptancedate. Received \receiveddate; in original form \originaldate}
\pubyear{2025}
\pubdate{\advpubdate}
\volume{537}
\makeatletter
\hypersetup{
    pdftitle={\@title},
    pdfauthor={Marnix J. Looijmans, Mike (Shengbo) Wang, Florian Beutler},
    pdfsubject={Monthly Notices of the Royal Astronomical Society, 537, 1, 21-34, 7-1-2025 (doi:10.1093/mnras/stae2786)},
    pdfkeywords={%
        methods: statistical;
        methods: data analysis;
        methods: numerical;
        large-scale structure of Universe
    }
}
\makeatother

\begin{document}
\label{firstpage}
\pagerange{\pageref{firstpage}--\pageref{lastpage}}

\maketitle

\begin{abstract}
    The determination of the covariance matrix and its inverse, the \emph{precision} matrix, is critical in the statistical analysis of cosmological measurements.
    The covariance matrix is typically estimated with a limited number of simulations at great computational cost before inversion into the precision matrix; therefore, it can be ill-conditioned and overly noisy when the sample size~\(\nsamp\) used for estimation is not much larger than the data vector dimension.
    In this work, we consider a class of methods known as \emph{shrinkage} estimation for the precision matrix, which combines an empirical estimate with a target that is either analytical or stochastic.
    These methods include linear and non-linear shrinkage applied to the covariance matrix (the latter represented by the so-called NERCOME estimator), and the direct linear shrinkage estimation of the precision matrix which we introduce in a cosmological setting.
    By performing Bayesian parameter inference and using metrics like matrix loss functions, the Kullback--Leibler divergence and the eigenvalue spectrum, we compare their performance against the standard sample estimator with varying sample size~\(\nsamp\).
    We have found the shrinkage estimators to significantly improve the posterior distribution at low \(\nsamp\), especially for the linear shrinkage estimators either inverted from the covariance matrix or applied directly to the precision matrix, with an empirical target constructed from the sample estimate.
    Our results are particularly relevant to the analyses of Stage-IV spectroscopic galaxy surveys such as the Dark Energy Spectroscopic Instrument and \textit{Euclid}, whose statistical power can be limited by the computational cost of obtaining an accurate precision matrix estimate. 
\end{abstract}


\Section{Introduction}
\label{sec:introduction}

\enlargethispage{-1.75\baselineskip}

With the arrival of Stage-IV spectroscopic galaxy surveys such as the Dark Energy Spectroscopic Instrument~(DESI)\footnote{\weblink{https://desi.lbl.gov}{desi.lbl.gov}} \citep{DESI2024,DESI2023} and \textit{Euclid}\footnote{\weblink{https://sci.esa.int/euclid}{sci.esa.int/euclid}, \weblink{https://euclid-ec.org}{euclid-ec.org}} \citep{Euclid2011}, we will be able to observe cosmic structure on scales larger than ever before.
At the same time, there has been significant development in perturbative models of non-linear clustering on smaller scales \citep[e.g.][]{Perko2016,dAmico2020,Semenaite2022}.
The wider dynamic range accessible to galaxy surveys poses a statistical challenge as the number of modes of cosmological fluctuations becomes very large, especially for the determination of the associated covariance matrix that is vital to most data analyses.

The covariance matrix character{\is}es the uncertainties and thus the constraining power of the analysis. 
Whilst it is sometimes possible to derive an analytical expression from perturbation theory \citep[e.g.][]{Grieb2016,Li2019,Mohammed2017, Sugiyama2020,Wadekar2020a}, including higher order corrections and accounting for systematic effects such as the survey window is far from straightforward \citep{Wadekar2020b}.
The alternative is an empirical approach whereby a sample estimate of the covariance matrix is obtained from a limited number of high-fidelity mock real{\is}ations.
The main challenge with this method is that the number of real{\is}ations depends on the dimension of the data vector, which can mean that thousands of mock catalogues are needed.
Because the evaluation of the likelihood function typically requires the inversion of the covariance matrix into the precision matrix, the covariance matrix estimate must not be singular, which is the case when the number of real{\is}ations is no greater than the data vector dimension.
Previous studies \citep{Taylor2013,Taylor2014,Sellentin2016} have also shown that using only a limited number of data real{\is}ations to estimate the covariance matrix induces noise to which the operation of matrix inversion is highly sensitive.
Moreover, since matrix inversion is a non-linear operation, an unbiased estimator of the covariance matrix does not generally result in an unbiased estimator of the precision matrix \citep{Anderson2003,Hartlap2006}.

\enlargethispage{-1.75\baselineskip}

To alleviate the issues associated with covariance matrix estimation, resampling methods such as jackknife and bootstrapping have been proposed \citep[e.g.][]{Friedrich2015,Escoffier2016,OConnell2019,Philcox2019,Mohammad2022}, which construct the covariance matrix from sub-samples of the observed data itself.
However, these sub-samples are obtained by dividing the original simulations into much smaller sub-volumes, which do not necessarily contain the largest modes of interest.
Other proposed methods include data vector dimensionality reduction with subspace projections \citep{Philcox2021}, covariance matrix denoising with convolutional neural networks \citep{deSanti2022} and covariance matrix model fitting with simulations \citep{Fumagalli2022}.
Precision matrices have also been directly studied in cosmological analysis; for example, \citet{Friedrich2018} uses a power series expansion of the precision matrix and estimate the error terms from simulations.

An alternative class of covariance estimation methods is \emph{shrinkage} estimation, which combines empirical estimates of the covariance matrix with a \emph{target} that is often (but not always) an analytical proxy.
The standard linear shrinkage method in the statistical literature \citep{Ledoit2003,Ledoit2004,Schafer2005} has been previously explored in a cosmological context by \citet{Pope2008}, and the NERCOME\footnote{Non-parametric Eigenvalue-Regular{\is}ed COvariance Matrix Estimator.} estimator from non-linear shrinkage methods \citep{Ledoit2012} has been first introduced by \citet{Joachimi2017} in the same context.
In a Bayesian setting, \citet{Hall2019} has derived a likelihood function conditioned on analytical and empirical covariance matrices assuming an inverse Wishart prior for the unknown true covariance matrix, which is analogous to the result obtained with the linear shrinkage estimator.
The Bayesian control variate method proposed by \citet{Chartier2022b} similarly combines different covariance matrix estimates from high- and low-fidelity simulations to reduce the statistical scatter, also assuming an inverse Wishart prior that plays a similar role to the shrinkage target.
To bridge the interpretations of parameter uncertainties in frequentist and Bayesian methods, \citet{Percival2022} instead advocates for a power-law form prior for the covariance matrix to approximately match the Bayesian credible interval and the frequentist confidence interval.
However, a limitation of these studies is that they primarily concern the covariance matrix estimate itself, which still needs to be inverted into the precision matrix estimate for likelihood analysis, and the effect of inversion is not yet fully understood.

In this work, we revisit the linear \citep{Pope2008} and non-linear \citep{Joachimi2017} covariance shrinkage estimation methods in a frequentist setting.
We will compute the precision matrix estimate after inversion using these two methods, and compare their performance with a third method: linear shrinkage directly applied to precision matrix estimates \citep{Bodnar2016}, introduced here in a cosmological context for the first time.

This paper is organ{\is}ed as follows: we introduce the different covariance and precision matrix estimators in section~\ref{sec:methods}, and apply them to a power spectrum analysis in section~\ref{sec:application}; we then compare their performance in section~\ref{sec:comparison}, and discuss our findings before concluding in section~\ref{sec:conclusion}.

For reference and clarity, our notation is listed in Table~\ref{tab:notation}.
\begin{table}
    \caption{Notation used in this work.}
    \begin{tabular*}{\columnwidth}{ll}
        \toprule
        \textbf{Symbol} & \textbf{Definition}  \\
        \midrule
        \(\diag*{\mat{A}}\) & Diagonal matrix consisting of the diagonal elements of \(\mat{A}\) \\
        \(\fnorm{\mat{A}}\) & Frobenius norm given by \(\left[\tr(\mat{A}\trs{\mat{A}})\right]^{1/2}\) \\
        \(\tnorm{\mat{A}}\) & Trace norm given by \(\tr\big[\left(\mat{A}\trs{\mat{A}}\right)^{1/2}]\) \\
        \(\intrange{a}{b}\) & Closed integer interval \\
        \midrule
        \(\nddim\) & Length of the data vector \\
        \(\nsamp\) & Number of data real{\is}ations \\
        \(\npara\) & Number of inferred parameters \\
        \(\datamat \in \real^{\nddim\times\nsamp}\) & Data matrix \\
        \(\covartrue \in \real^{\nddim\times\nddim}\) & True covariance matrix \\
        \(\prectrue \in \real^{\nddim\times\nddim}\) & True precision matrix given by \(\covartrue^{-1}\) \\
        \midrule
        \(\covarsamp\) & Sample estimator of \(\covartrue\) \\
        \(\precsamp\) & Sample estimator of \(\prectrue\) \\
        \(\covartar\) & Covariance matrix target \\
        \(\prectar\) & Precision matrix target \\
        \(\estm{\covartrue}_\mathrm{LS}\) & Linear shrinkage estimator of \(\covartrue\) \\
        \(\estm{\prectrue}_\mathrm{LS}\) & \(\estm{\covartrue}_\mathrm{LS}^{-1}\), inverted covariance shrinkage estimator of \(\prectrue\) \\
        \(\estm{\covartrue}_\mathrm{NLS}\) & Non-linear shrinkage (NERCOME) estimator of \(\covartrue\) \\
        \(\estm{\prectrue}_\mathrm{NLS}\) & \(\estm{\covartrue}_\mathrm{NLS}^{-1}\), NERCOME estimator of \(\prectrue\) \\
        \(\precsamp_{\mkskip\mathrm{LS}}\) & Direct linear shrinkage estimator of \(\prectrue\) \\
        \(\lambda\) & Linear-shrinkage intensity parameter \\
        \(\alpha, \beta\) & Linear-shrinkage mixing parameters \\
        \(s\) & NERCOME split parameter \\
        \bottomrule
    \end{tabular*}
    \label{tab:notation}
\end{table}
For reproducibility and reference, the implementation of the different estimation methods considered in this work can be found in our public repository online (see \hyperlink{data_availability}{Data Availability} for details).

\Section{Estimation Methods}
\label{sec:methods}

In this section, we describe the different precision matrix estimation methods to be compared.
Most of them invert a covariance matrix estimate at the final stage, but the new method we adopt involves matrix inversion only as an intermediate step. The implications of covariance matrix inversion will be discussed.

\subsection{Sample estimation}
\label{subsec:sample_estimation}

Suppose we have a data matrix~\(\datamat \in \real^{\nddim \times \nsamp}\) consisting of \(\nsamp\) independent real{\is}ations of a random vector~\(\vec{Y}\) of length~\(\nddim\), which has an underlying covariance matrix~\(\covartrue \in \real^{\nddim \times \nddim}\) that is a priori unknown.
Without loss of generality, we assume the data vectors are mean-subtracted.
The standard sample estimator of the covariance matrix is then given by
\begin{equation}
    \label{eq:sample_cov}
    \covarsamp = \frac{1}{\nsamp - 1} \datamat \trs{\datamat} \,.
\end{equation}
It is unbiased with expectation~\(\Exp[\covarsamp] = \covartrue\).

In cosmological analysis, one typically assumes that the measurement vector~\(\vec{Y} \sim \func{\mathcal{N}}(\bm{\mu}, \covartrue)\) follows the multivariate normal distribution, as is the case for compressed summary statistics under the central limit theorem such as the band power spectrum from averaging over a large number of clustering modes.
The likelihood function of model parameters~\(\bm{\theta}\) then takes the form
\begin{equation}
    \label{eq:likelihood}
    \Like\big(\bm{\theta}) \propto (\det\covartrue)^{-\ifrac{1}{2}} \exp\left[- \frac{1}{2} \trs{(\vec{Y} - \bm{\mu})} \covartrue^{-1} (\vec{Y} - \bm{\mu})\right] \,,
\end{equation}
where \(\bm{\mu} \equiv \func{\bm{\mu}}(\bm{\theta})\) is the mean/model vector for the measurements, and the underlying covariance matrix~\(\covartrue \equiv \func{\covartrue}(\bm{\theta})\) may also depend on \(\bm{\theta}\).
Here it is the precision matrix~\(\prectrue \equiv \covartrue^{-1}\) that appears instead of the covariance matrix, so one may wish to estimate the precision matrix.
From equation~\eqref{eq:sample_cov}, we obtain the sample precision matrix estimator
\begin{equation}
    \label{eq:sample_pre}
    \precsamp = \hfac \covarsamp^{-1} \,,
\end{equation}
where the prefactor
\begin{equation}
    \label{eq:Hartlap_factor}
    \hfac = \frac{\nsamp - \nddim - 2}{\nsamp - 1}
\end{equation}
corrects for the multiplicative bias resulting from the inversion of the covariance matrix estimator~\(\covarsamp\), which is a non-linear operation, in the case of multivariate normal data.
It is immediately clear that the number of data real{\is}ations must satisfy \(\nsamp > \nddim + 2\), and in the limit \(\nsamp \to \infty\), \(\covarsamp^{-1}\) represents an asymptotically unbiased precision matrix estimator in itself.

The prefactor~\(\hfac\) here is commonly known as the `Hartlap factor' in cosmological analyses \citep{Hartlap2006}, and is derived from the inverse Wishart distribution for \(\covarsamp^{-1}\) when the data is multivariate normal \citep{Anderson2003}.
In this work, we do not consider the impact of non-Gaussianity in the data vector and associated covariance matrix, but instead refer the reader to e.g. \citet{Sellentin2018} and \citet{Blot2016,Blot2019} for more detailed discussion.

\subsection{Linear shrinkage}
\label{subsec:inverted_linear_shrinkage}

The \emph{linear shrinkage} (LS) method uses a convex combination of an empirical covariance matrix estimate~\(\estm{\covartrue}\) and a \emph{target} matrix~\(\covartar\) to construct a new covariance matrix estimator \citep{Ledoit2003,Ledoit2004},
\begin{equation}
    \label{eq:shrinkage_cov}
    \estm{\covartrue}_\mathrm{LS} = (1 - \lambda) \estm{\covartrue} + \lambda \covartar \,,
\end{equation}
where \(\lambda \in [0, 1]\) is the shrinkage \emph{intensity}.
The empirical estimate~\(\estm{\covartrue}\) is commonly chosen to be the \emph{unbiased} sample covariance matrix estimate~\(\covarsamp\) in equation~\eqref{eq:sample_cov}, which we shall assume for the rest of this work; the target~\(\covartar\) is often, but not always, an analytical quantity.

The idea behind shrinkage methods is that the target matrix has a smaller variance (possibly zero) than the empirical estimate, but may be otherwise biased, and one aims to find the optimal trade-off between bias and variance.
Following \citet{Pope2008}, we first define
\begin{equation}
    \mean{W}_{ij} = \frac{1}{\nsamp} \sum_{k=1}^{\nsamp} W_{ij}^{(k)}
    \quad \text{with} \quad 
    W_{ij}^{(k)} = x_i^{(k)} x_j^{(k)} \,,
\end{equation}
where \(x_i^{(k)}\) is the \((i, k)\)-th element of \(\datamat\).
The elements of the sample covariance matrix are then given by
\begin{equation}
    \est{S}_{ij} = \frac{\nsamp}{\nsamp-1} \mean{W}_{ij} \,,
\end{equation}
whose covariance can be estimated by
\begin{equation}
    \eCov(S_{ij}, S_{lm}) = \frac{\nsamp}{(\nsamp-1)^3} \sum_{k=1}^{\nsamp} \left(W_{ij}^{(k)} - \mean{W}_{ij}\right) \left(W_{lm}^{(k)} - \mean{W}_{lm}\right) \,.
\end{equation}
\citet{Ledoit2003} and \citet{Schafer2005} have shown that the optimal intensity parameter is estimated by
\begin{equation}
    \label{eq:optimal_intensity}
    \opt{\estm{\lambda}} = \frac{\sum_{i,j} \left[\eVar(S_{ij}) - \eCov(S_{ij}, T_{ij})\right]}{\sum_{i,j} \left({S}_{ij} - T_{ij}\right)^2} \,,
\end{equation}
where \(\eVar(S_{ij}) \equiv \eCov(S_{ij}, S_{ij})\).
If \(\opt{\estm{\lambda}}\) is found to be negative, it is then clipped to \(\opt{\estm{\lambda}} = 0\); if it is greater than 1, it is clipped to \(\opt{\estm{\lambda}} = 1\).

Finally, an estimate of the precision matrix is obtained by inversion, \(\estm{\prectrue}_\mathrm{LS} = \estm{\covartrue}_\mathrm{LS}^{-1}\).
Since there is no general analytical form for the probability distribution of \(\estm{\covartrue}_\mathrm{LS}\), unlike for the sample estimator~\(\covarsamp\) which follows the Wishart distribution for multivariate normal data \citep{Anderson2003}, we do not include an analogous Hartlap factor here \citep{Torre2013}.

\subsection{Non-linear shrinkage}
\label{subsec:non-linear_shrinkage}

The linear shrinkage estimator is in fact a first-order approximation to a more general class of \emph{non-linear shrinkage} (NLS) estimators that attempt to shrink the dynamic range of the eigenvalues of the covariance matrix estimate and thus mitigate ill-conditioning when the data vector dimension becomes large \citep{Ledoit2012}. 
In this work, we consider in particular the \emph{NERCOME estimator} of the covariance matrix proposed by \citet{Ledoit2012}, \cite{Abadir2014} and \citet{Lam2016}, which has been previously applied by \citet{Joachimi2017} and \citet{GouyouBeauchamps2023} in a cosmological setting.

The NERCOME estimator is constructed from a set of data real{\is}ations as follows:
\begin{enumerate}
    \item Without necessarily preserving their order, the columns of the data matrix~\(\datamat\) are split into a \(\nddim \times s\) matrix~\(\datamat_1\) and a \(\nddim \times (\nsamp - s)\) matrix~\(\datamat_2\) for a given value of the split parameter~\(s \in \intrange{2}{\nsamp - 2}\);
    \item Two sample estimates of the covariance matrix, denoted \(\covarsamp_1\) and \(\covarsamp_2\), are obtained from \(\datamat_1\) and \(\datamat_2\) respectively;
    \item The sample estimates are diagonal{\is}ed, \(\covarsamp_a = \mat{U}_a \mat{D}_a \trs{\mat{U}}_a\) with \(a \in \{1, 2\}\), where \(\mat{D}_a\) is the diagonal matrix of eigenvalues and \(\mat{U}_a\) is the orthogonal matrix of eigenvectors;
    \item The quantity~\(\est{\mat{Z}} = \mat{U}_1 \diag(\trs{\mat{U}}_1 \covarsamp_2 \mat{U}_1) \trs{\mat{U}}_1\) is computed for a given split at the chosen split parameter value~\(s\);
    \item Over all \(\binom{\nsamp}{s}\) possible choices of the split into \(\datamat_{1,2}\) for the chosen value of \(s\), the average of \(\est{\mat{Z}}\) is calculated to be \(\mean{\mat{Z}}(s)\), and correspondingly the sample estimates~\(\covarsamp_2\) are averaged to obtain \(\mean{\covarsamp}_2(s)\);
    \item The \emph{distance} function
    \begin{equation}
        \label{eq:nercome_distance_func}
        Q(s) = \fnorm[\big]{\mean{\mat{Z}}(s) - \mean{\covarsamp}_2(s)}
    \end{equation}
    (or equivalently \(Q^2\)) is minim{\is}ed to find an optimal split value \(\opt{s}\).
\end{enumerate}
The NERCOME estimator of the covariance matrix is then defined as \(\estm{\covartrue}_\mathrm{NLS} = \mean{\mat{Z}}(\opt{s})\).
As before, the precision matrix estimator is given by its inverse, \(\estm{\prectrue}_\mathrm{NLS} = \mean{\mat{Z}}(\opt{s})^{-1}\), and we do not include a Hartlap-like factor.

\subsection{Direct linear shrinkage for the precision matrix}
\label{subsec:direct_linear_shrinkage}

Instead of inverting the linear shrinkage estimator of the covariance matrix, one can apply shrinkage directly to an empirical estimate and a target for the precision matrix.
Here we consider the inverse of the sample covariance matrix estimate, \(\covarsamp^{-1}\), as the empirical estimate, with some suitable precision matrix target denoted by \(\prectar\) to give the direct linear shrinkage estimator of the precision matrix:
\begin{equation}
    \label{eq:shrinkage_pre}
    \precsamp_{\mkskip\mathrm{LS}} = \alpha \covarsamp^{-1} + \beta \mkern1mu \prectar \,.
\end{equation}
Here \(\alpha\) and \(\beta\) are the \emph{mixing parameters}, and \(\nsamp > \nddim\) must be satisfied to ensure that \(\covarsamp\) is non-singular; in addition, the target \(\prectar\) is required to satisfy the condition \(\sup_\nddim \nddim^{-1} \fnorm{\prectar}^2 \leqslant a\) for some constant~\(a > 0\), so that \(\beta\) remains bounded as \(\nddim \to \infty\).
\citet{Bodnar2016} have shown that to minim{\is}e the Frobenius loss \(\fnorm{\precsamp_{\mkskip\mathrm{LS}} - \prectrue}\), the optimal mixing parameters can be estimated by
\begin{subequations}
\label{eq:optimal_mixing}
\begin{align}
    \label{eq:optimal_alpha}
    \opt{\estm{\alpha}} &= 1 - \frac{\nddim}{\nsamp} - \frac{\nsamp^{-1} \tnorm{\covarsamp^{-1}}^2 \mkskip \fnorm{\prectar}^2}{\fnorm{\covarsamp^{-1}}^2 \mkskip \fnorm{\prectar}^2 - \left[\tr(\covarsamp^{-1} \prectar)\right]^2} \,, \\
    \opt{\estm{\beta}} &= \left(1 - \frac{\nddim}{\nsamp} - \opt{\estm{\alpha}}\right) \frac{\tr\big(\covarsamp^{-1} \prectar)}{\fnorm{\prectar}^2} \,,
\end{align}    
\end{subequations}
which converge almost surely to their respective asymptotic values as \(\nsamp \to \infty\) with \(\ifrac{\nddim}{\nsamp} \to c \in (0, 1)\) for some fixed constant~\(c\), resulting in a consistent precision matrix estimator.

We note here that the target~\(\prectar\) needs to be chosen carefully: if it is too close to \(\covarsamp^{-1}\), negative \(\opt{\estm{\alpha}}\) values can occur and render the estimate~\(\precsamp_{\mkskip\mathrm{LS}}\) no longer positive semi-definite; indeed, if \(\prectar = \covarsamp^{-1}\) is chosen, the denominator in equation~\eqref{eq:optimal_alpha} vanishes.
This situation is not unique to the estimator considered here, and can happen to other shrinkage-like estimators, e.g. the CARPool estimator in \citet{Chartier2022a}.
In any case, the positive semi-definiteness of a covariance/precision matrix estimate can be easily checked, and the specific target matrix replaced with another.

\Section{Application to Power Spectrum Analysis}
\label{sec:application}

To demonstrate the performance of the estimation methods reviewed in the previous section, we apply them to a galaxy clustering power spectrum analysis of the Baryon Oscillations Spectroscopic Survey (BOSS) Data Release 12 (DR12) catalogues \citep{Alam2015,Reid2016} for the North Galactic Cap (NGC) in the redshift bin at \(z = 0.38\), with the accompanying \num{2048} \codename{patchy} mock catalogues \citep{Kitaura2016} for covariance matrix estimation.
The data set we use includes measurements of the clustering power spectrum monopole~\(P_0\) and quadrupole~\(P_2\) over the wavenumber range \(k \in [0.01, 0.1] \si{\h\per\mega\parsec}\) in bins of width \(\Delta{k} = \SI{0.01}{\h\per\mega\parsec}\), which results in concatenated data vectors of length \(\nddim = 18\).\footnote{\label{fnote:dataset}Full data set including the transformation matrices are available at \weblink{https://fbeutler.github.io/hub/deconv_paper}{fbeutler.github.io/hub/deconv\_paper}.}

Given the wavenumber range considered in this work, we adopt a linear redshift-space power spectrum model \citep{Kaiser1987} computed from \citet{Eisenstein1998}, with the fiducial cosmology taken from the Planck~2015 results \citep{Planck2015}.
The power spectrum model for the monopole, quadrupole and hexadecapole is given by
\begin{subequations}
\label{eq:kaiser_model}
\begin{align}
    P_0(k) &= \left(b^2 + \frac{2}{3} bf + \frac{1}{5} f^2\right) P_\mathrm{m}(k) \,, \\
    P_2(k) &= \left(\frac{4}{3}bf + \frac{4}{7}f^2 \right) P_\mathrm{m}(k) \,, \\
    P_4(k) &= \frac{8}{35} f^2 P_\mathrm{m}(k) \,,
\end{align}
\end{subequations}
where \(P_\mathrm{m}\) is the linear matter power spectrum, \(b\) is the linear bias and \(f\) is the linear growth rate.
The linear bias~\(b\) and growth rate~\(f\) will be the parameters of interest when we perform statistical inference in section~\ref{subsec:parameter_inference}.
In addition, we include the wide-angle and survey window effects as matrices~\(\mat{W}\) and \(\mat{M}\) respectively \citep{Beutler2021},\footnoteref{fnote:dataset} which transform the power spectrum model vector~\(\vec{P}_\mathrm{model} \mapsto \mat{W}\mat{M}\vec{P}_\mathrm{model}\) evaluated in the same bins as for the measurements.

To assess the performance of the different estimators with varying sample size, we have created a superset of \num{2040} power spectrum measurements from the \num{2048} mock catalogues available.
This superset is divided into \num{85} subsets of \num{24} mock measurements (i.e. \num{85} data matrices~\(\datamat \in \real^{18\times24}\)), or \num{68} subsets of \num{30} mock measurements each (i.e. \num{68} data matrices~\(\datamat \in \real^{18\times30}\)), so that we can test covariance and precision matrix estimation with \(\nsamp = \num{24}\) and \(\nsamp = \num{30}\) sample sizes respectively.

Here we remark that the cosmological sample above is chosen as it is publicly available with a large number of companion mock catalogues as well as the window convolution matrix.
The wavenumber range considered is such that a simple linear Kaiser model suffices without the need for computationally expensive non-linear modelling in this proof-of-concept study, while the inclusion of survey window effects means that the structure of the covariance/precision matrix is not purely diagonal.
An actual Stage-IV survey analysis will exploit non-linear clustering out to much smaller scales with more complicated non-Gaussian clustering statistics and covariance matrices \citep[see e.g.][]{Colavincenzo2019,Karacayli2024,Brown2024,Ajani2023}, and it will be interesting to repeat and expand upon the analysis presented in this work.
Nevertheless, how well the different precision matrix estimation techniques perform, especially with a very limited number of mock real{\is}ations, will still be informative for Stage-IV survey analyses involving a much higher dimensional data vector.

\subsection{Sample estimates}
\label{subsec:sample_estimates}

For each subset of the mock measurements, we compute a sample covariance matrix estimate~\(\covarsamp\) using equation~\eqref{eq:sample_cov}, which is then inverted into a sample precision matrix estimate~\(\precsamp\) with the Hartlap factor~\(\hfac\) being \num{0.174} for \(\nsamp = \num{24}\) and \num{0.344} for \(\nsamp = \num{30}\) respectively.

As a \emph{reference} covariance matrix, we also compute the sample covariance matrix estimate~\(\covarref\) from all \num{2048} mock measurements, which is used to benchmark all other covariance matrix estimators.
The corresponding sample precision matrix estimate~\(\precref\) is used as a reference precision matrix.
This has a Hartlap factor~\(\hfac = \num{0.991} \simeq 1\) as \(\nsamp = \num{2048} \gg \nddim = 18\), and we regard it as a good proxy for the unknown true precision matrix~\(\prectrue\).
The absolute values of the reference covariance and precision matrices are shown in Fig.~\ref{fig:ref_cov_prec}, which exhibits their mostly block-diagonal structure with off-diagonal correlations mainly induced by the survey window effect.
\begin{figure}
    \includegraphics[width=\columnwidth]{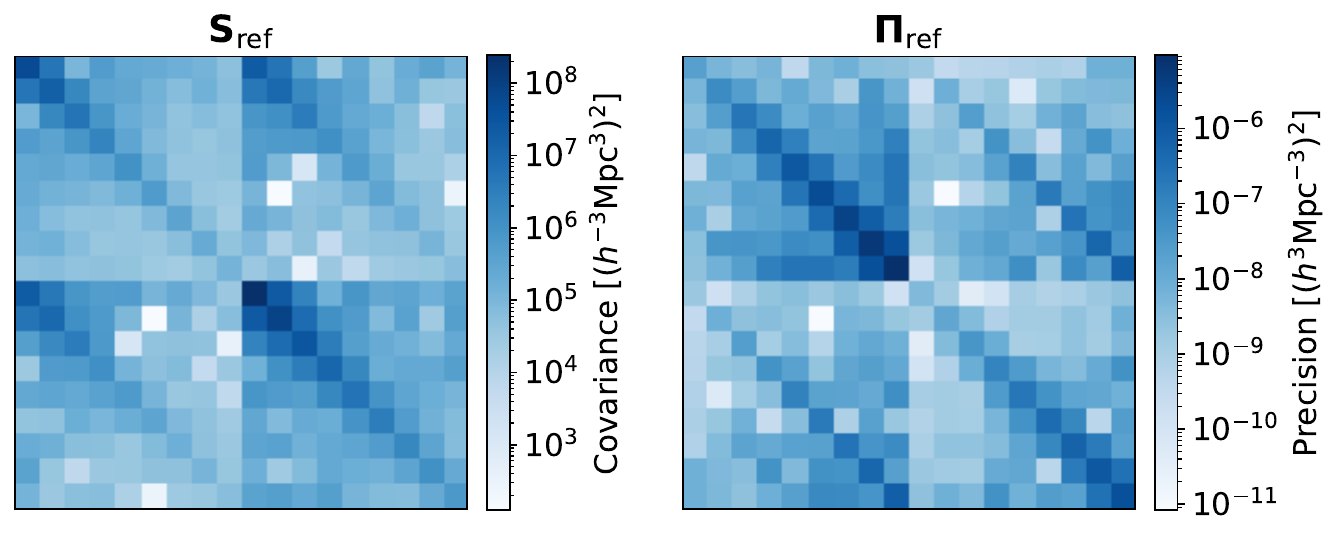}
    \caption{
        Colour maps for the absolute values of the reference covariance matrix~\(\covarref\) (\textit{left panel}) and the reference precision matrix~\(\precref\) (\textit{right panel}).
    }
    \label{fig:ref_cov_prec}
\end{figure}

\subsection{Inverted covariance shrinkage estimates}
\label{subsec:inverted_linear_shrinkage_estimates}

For the linear shrinkage estimator~\eqref{eq:shrinkage_cov} of the covariance matrix, we consider two possibilities for the target matrix~\(\covartar\).
The first choice, proposed by \citet{Schafer2005}, is simply the diagonal of the sample covariance matrix estimate:
\begin{equation}
    \label{eq:diag_cov_target}
    \covartar^{(1)} = \diag*{\covarsamp} \,.
\end{equation}
In this case, only summands with \(i \neq j\) contribute to the numerator and denominator of expression~\eqref{eq:optimal_intensity}.
The second choice of target, following \citet{Hamilton2006} and \citet{Pope2008}, is given by
\begin{equation}
    \label{eq:analy_cov_target}
    \big(\covartar^{(2)}\big)_{ij} = \frac{2}{N_i} P_\mathrm{fid}(k_i)^2 \delta^\mathrm{(K)}_{ij} \,,
\end{equation}
where \(P_\mathrm{fid}\) is a fiducial power spectrum model for which we set \(b = 2\) and \(f = 0.7\) and transform using the wide-angle and survey window matrices, and \(N_i\) is the number of modes in the bin at effective wavenumber \(k_i\).
Since \(\covartar^{(2)}\) is deterministic, \(\eCov(S_{ij}, T_{ij})\) identically vanishes in equation~\eqref{eq:optimal_intensity}.

For both choices of the target matrix, we compute the linear shrinkage estimate~\(\estm{\covartrue}_\mathrm{LS}\) of the covariance matrix from each subset of mock measurements; this gives \num{85} estimates with \(\nsamp = \num{24}\) and \num{68} estimates with \(\nsamp = \num{30}\).
The distributions of the optimal shrinkage intensity parameter~\(\opt{\estm{\lambda}}\) for these two target choices are compared in Fig.~\ref{fig:dist_optimal_intensity}.
\begin{figure}
    \includegraphics[width=0.90\columnwidth]{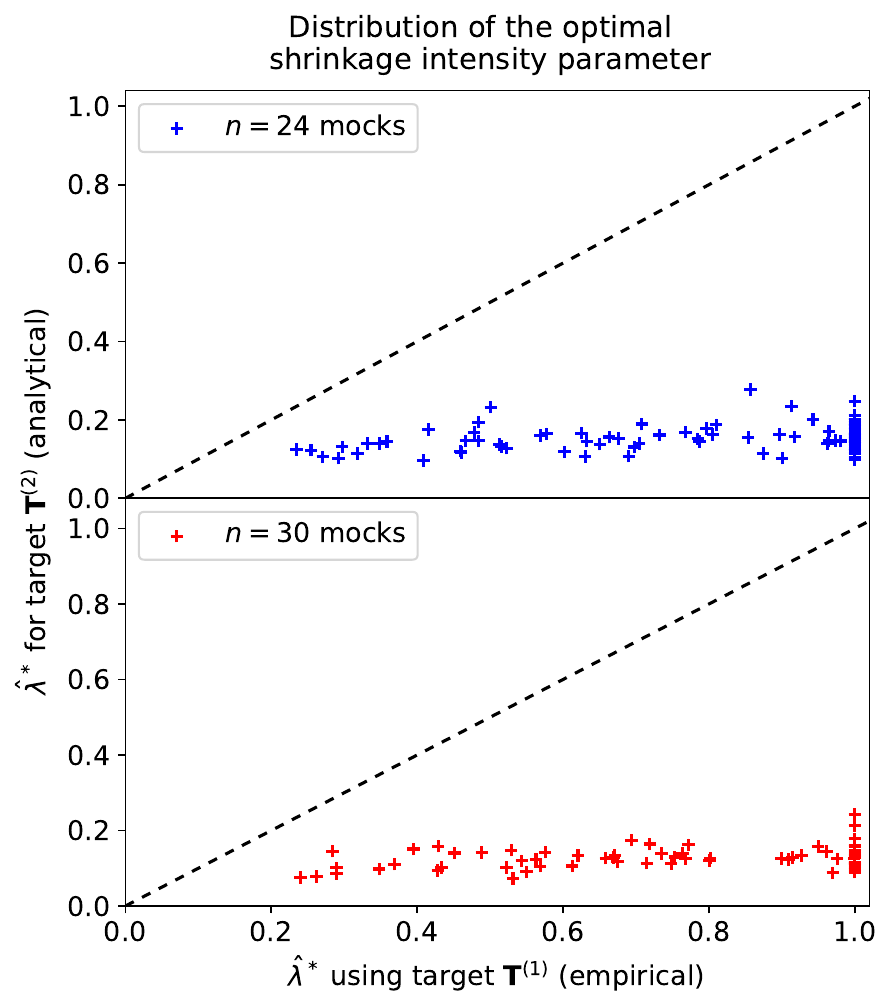}
    \caption{
        Distributions of the optimal shrinkage intensity~\(\opt{\estm{\lambda}}\) for target matrix choices \(\covartar^{(1)}\) versus \(\covartar^{(2)}\).
        The \textit{top panel} shows the \num{85} intensity parameters computed from subsets of \(\nsamp = 24\) mock measurements; the \textit{bottom panel} shows the \num{68} intensity parameters computed for \(\nsamp = 30\).
        The diagonal dashed lines show the line of equality for \(\opt{\estm{\lambda}}\) between the two target choices.
    }
    \label{fig:dist_optimal_intensity}
\end{figure}

We have found that the shrinkage intensity~\(\opt{\estm{\lambda}}\) with target~\(\covartar^{(1)}\) is equal to unity for \num{31} out of \num{85} data matrices with \(\nsamp = \num{24}\) and for \num{21} out of \num{68} data matrices with \(\nsamp = \num{30}\).
This indicates that in these cases, the linear shrinkage estimate in equation~\eqref{eq:shrinkage_cov} is dominated by the target with no contribution from the sample estimate.
Moreover, the spread of \(\opt{\estm{\lambda}}\) for target~\(\covartar^{(1)}\) (from \num{0.2} to \num{1.0}) is much larger than for target~\(\covartar^{(2)}\), because the empirical target~\(\covartar^{(1)}\) is itself obtained from the noisy sample covariance matrix~\(\covarsamp\) estimated with only \(\nsamp = \num{24}\) or \num{30} mock measurements; this means the linear shrinkage estimate may be more adaptive to the scatter in the sample estimate~\(\covarsamp\).
In contrast, with target~\(\covartar^{(2)}\), the shrinkage intensity~\(\opt{\estm{\lambda}}\) favours smaller values, which means that the target tends to contribute less to the shrinkage estimate than the sample estimate does.
If the reference covariance matrix~\(\covarref\) is used as \(\covarsamp\) in the linear shrinkage estimator, the optimal intensity parameter is \(\opt{\estm{\lambda}} \approx \num{0.028}\) for target~\(\covartar^{(1)}\) and \(\opt{\estm{\lambda}} \approx \num{0.002}\) for target~\(\covartar^{(2)}\).
In both cases, the linear shrinkage estimator is almost fully determined by the sample estimate of the covariance matrix.
This is expected since our reference covariance matrix~\(\covarref\) is supposed to closely approximate the true covariance matrix~\(\covartrue\).

To obtain the corresponding precision matrix estimates~\(\estm{\prectrue}_\mathrm{LS}\), the aforementioned linear-shrinkage covariance matrix estimates are simply inverted.

\subsection{NERCOME estimates}
\label{subsec:non-linear_shrinkage_estimates}

We follow the steps detailed in section~\ref{subsec:non-linear_shrinkage} to compute \num{85} NERCOME covariance matrix estimates~\(\estm{\covartrue}_\mathrm{NLS}\) from the data matrices~\(\datamat \in \real^{18\times24}\) and \num{68} estimates from the data matrices~\(\datamat \in \real^{18\times30}\).
For reference, we also compute a single NERCOME estimate using all \num{2048} mock measurements.
As before, the corresponding precision matrix estimates~\(\estm{\prectrue}_\mathrm{NLS}\) are obtained by matrix inversion.

A caveat here is that since \(\binom{\nsamp}{s}\) can be very large for some values of \(s\), we only randomly draw \(N_\mathrm{draw} = \min\left\{\binom{\nsamp}{s}, \num{1000}\right\}\) splits of the data matrix~\(\datamat\) at step~(v) in section~\ref{subsec:non-linear_shrinkage} to compute \(\mean{\mat{Z}}(s)\) and \(\mean{\covarsamp}_2(s)\) \citep{Joachimi2017}.

To minim{\is}e the distance function~\(Q(s)\) in equation~\eqref{eq:nercome_distance_func}, we evaluate it at each point in the integer interval~\(s \in \intrange{2}{\nsamp - 2}\), except when \(\nsamp = 2048\), in which case we evaluate \(Q(s)\) only at \(s \in \left\{0.1\nsamp, 0.15\nsamp, 0.2\nsamp, \dots, 0.9\nsamp, 2\sqrt{\nsamp}, \nsamp - 1.5\sqrt{\nsamp},\, \nsamp - 2.5\sqrt{\nsamp}\right\}\), as proposed by \citet{Joachimi2017} and \citet{Lam2016}.

The distribution of the optimal split parameters \(\opt{s}\) for all NERCOME estimates computed with \(\nsamp = \num{24}\) and \(\nsamp = \num{30}\) is shown in Fig.~\ref{fig:dist_optimal_split}.
\begin{figure}
    \includegraphics[width=0.90\columnwidth]{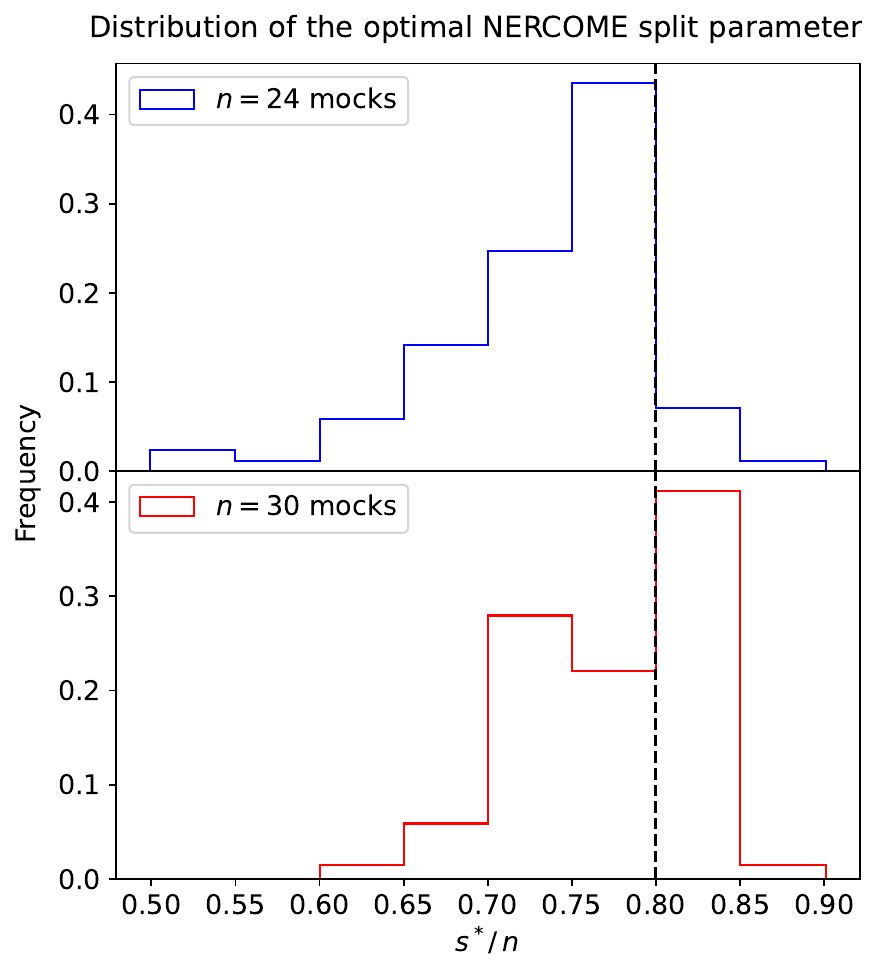}
    \caption{
        Normal{\is}ed histogram showing the distribution of the optimal split parameter~\(\opt{s}\) that minim{\is}es the distance function~\(Q(s)\) (equation~\ref{eq:nercome_distance_func}) as a ratio~\(\ifrac{s}{\nsamp}\) to the number of data real{\is}ations.
        The \textit{top panel} corresponds to \num{85} NERCOME covariance matrix estimates computed from subsets of \(\nsamp = \num{24}\) mock measurements; the \textit{bottom panel} corresponds to \num{68} estimates computed for \(\nsamp = \num{30}\).
        The vertical dashed line shows the optimal split parameter for \(\nsamp = \num{2048}\) mock measurements.
    }
    \label{fig:dist_optimal_split}
\end{figure}
The split-parameter distribution indicates that \(\ifrac{\opt{s}}{\nsamp} \simeq 0.8\) is favoured.
For reference, the split parameter for the NERCOME estimate computed using all \num{2048} mocks is \(\ifrac{\opt{s}}{\nsamp} = \ifrac{\num{1638}}{\num{2048}} \simeq \num{0.80}\).

\subsection{Direct precision shrinkage estimates}
\label{subsec:direct_linear_shrinkage_estimates}

Finally, for the direct linear shrinkage estimator~\eqref{eq:shrinkage_pre} of the precision matrix, we consider two possibilities for the target matrix~\(\prectar\).
The first choice is the matrix consisting of eigenvalues of the inverse sample covariance matrix estimate on the diagonal, i.e.
\begin{equation}
    \big(\prectar^{(1)}\big)_{ij} = e_i \delta^\mathrm{(K)}_{ij},
\end{equation}
where \(\{e_i\}_{i=1}^{\nddim}\) are the eigenvalues of \(\covarsamp^{-1}\) in ascending order.
The second choice is to invert the analytical covariance matrix target from equation~\eqref{eq:analy_cov_target},
\begin{equation}
    \prectar^{(2)} = \big(\covartar^{(2)}\big)^{-1}.
\end{equation}
We then compute two linear shrinkage estimates~\(\precsamp_{\mkskip\mathrm{LS}}\) of the precision matrix, one per target, for each subset of mock measurements with \(\nsamp = \num{24}\) or \(\nsamp = \num{30}\) using equation~\eqref{eq:shrinkage_pre}.

The estimated optimal shrinkage mixing parameters~\(\opt{\estm{\alpha}}\) and \(\opt{\estm{\beta}}\) are computed using equations~\eqref{eq:optimal_mixing} and shown in Fig.~\ref{fig:dist_optimal_mixing}.
\begin{figure*}
    \includegraphics[width=0.7\textwidth]{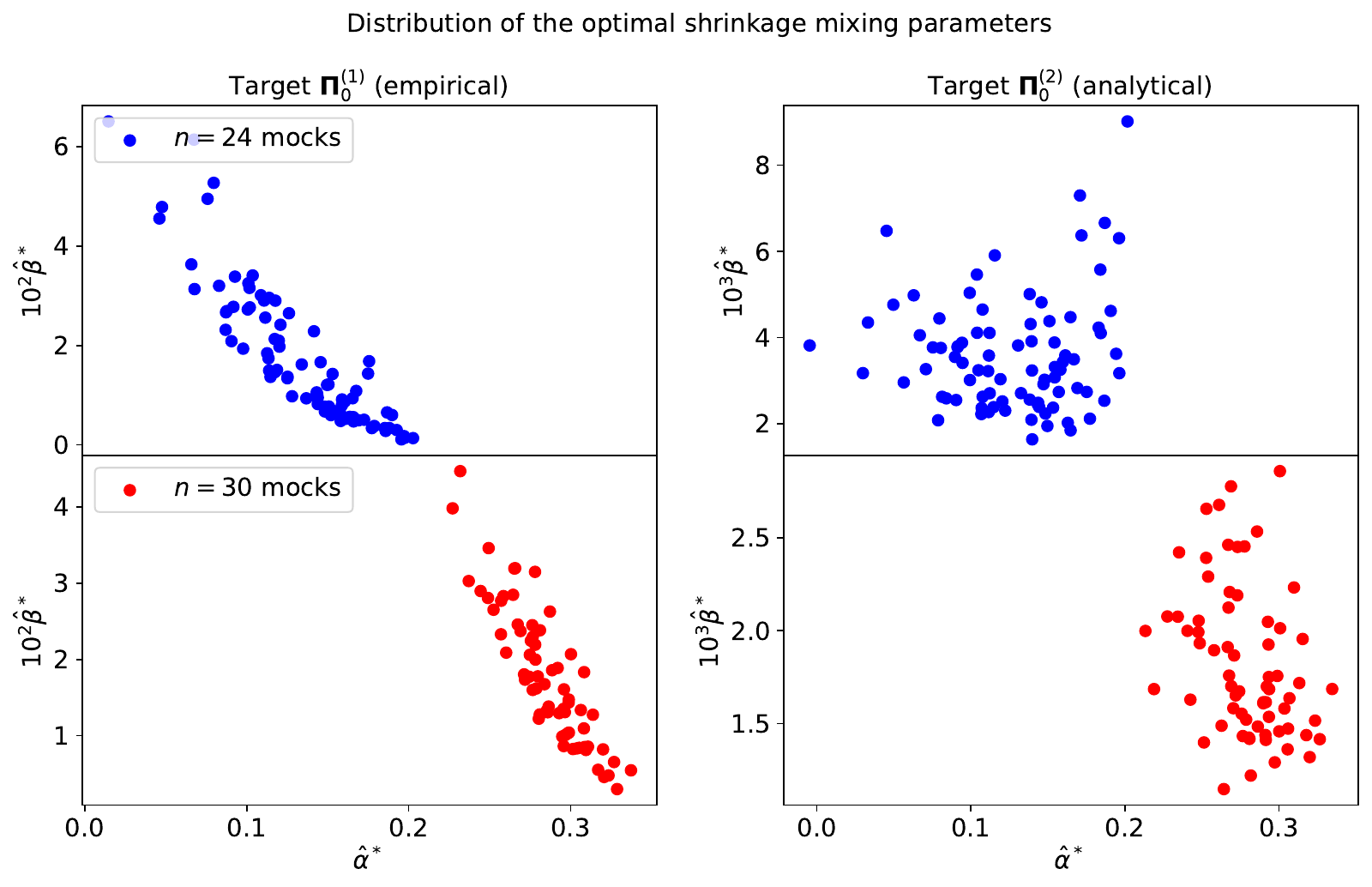}
    \caption{
        Distributions of the optimal linear-shrinkage mixing parameters~\(\opt{\estm{\alpha}}\) and \(\opt{\estm{\beta}}\) for two choices of the target matrix~\(\prectar\) used in the direct precision matrix estimate~\(\estm{\precsamp}_\mathrm{LS}\).
        Each \textit{top panel} shows the \num{85} mixing parameters computed for subsets of \(\nsamp = \num{24}\) mock measurements, and each \textit{bottom panel} shows the \num{68} mixing parameters computed for \(\nsamp = \num{30}\).
        Note the different scales on the vertical axes.
    }
    \label{fig:dist_optimal_mixing}
\end{figure*}
We note that \(\opt{\estm{\alpha}}\) and \(\opt{\estm{\beta}}\) have an approximate linear relationship when \(\prectar^{(1)}\) is used as target.
When we use \(\prectar^{(2)}\) as target, the linear relationship disappears.
For target \(\prectar^{(2)}\), we have noted that \(\opt{\estm{\alpha}}\) is negative for one out of the \num{85} direct precision shrinkage estimates with \(\nsamp = \num{24}\), indicating that the signs of all the entries in \(\covarsamp^{-1}\) are flipped for this estimate.
The mixing parameter \(\opt{\estm{\alpha}}\) is never negative for any other number of mock measurements \(\nsamp\) or target~\(\prectar^{(1)}\).
Moreover, it can be seen from the distribution of mixing parameters that \(\opt{\estm{\alpha}}\) becomes larger when a larger number of mock measurements is used.
This indicates that the inverse sample covariance matrix estimate $\covarsamp^{-1}$ has greater weight in shrinkage with larger numbers of mock measurements, as one would expect.
Finally, the values of \(\opt{\estm{\beta}}\) for the target~\(\prectar^{(2)}\) are smaller by an order of magnitude than those for the  the target~\(\prectar^{(1)}\), implying that the analytical target is generally less favoured than the empirical target.

For reference, we have also listed in Table~\ref{tab:shrink_mix_2048} the optimal shrinkage mixing parameters when \(\covarref\) is used for the precision matrix estimate in equation~\eqref{eq:shrinkage_pre}.
As expected, the \(\opt{\estm{\alpha}}\) values are close to 1 and the \(\opt{\estm{\beta}}\) values are close to 0.
\begin{table}
    \caption{
        Linear-shrinkage mixing parameters for the direct precision matrix estimate computed with \(\covarsamp = \covarref\) in equation~\eqref{eq:shrinkage_pre} for different target matrices.
    }
    \centering
    \begin{tabular*}{\columnwidth}{@{\hspace{2em}}l@{\extracolsep{\fill}}l@{\extracolsep{3em}}l@{\hspace{2em}}}
        \toprule
        {Target} & \(\opt{\estm{\alpha}}\) & \(\opt{\estm{\beta}}\) \\
        \midrule
        \(\prectar^{(1)}\) (empirical) & \num{0.989} & \num{4.71e-4} \\[2pt]
        \(\prectar^{(2)}\) (analytical) & \num{0.989} & \num{1.39e-5} \\
        \bottomrule
    \end{tabular*}
    \label{tab:shrink_mix_2048}
\end{table}

\Section{Performance Comparison}
\label{sec:comparison}

We will next compare the performance of seven types of precision matrix estimate constructed from the previous section for application to the BOSS DR12 power spectrum analysis, which are listed below for clarity and completeness:
\begin{enumerate}
    \item the \textbf{sample estimate}~\(\precsamp\) introduced in section~\ref{subsec:sample_estimation} and computed in section~\ref{subsec:sample_estimates};
    \item the inverted covariance linear-shrinkage estimate, \(\estm{\prectrue}_\mathrm{LS}\), introduced in section~\ref{subsec:inverted_linear_shrinkage} with the empirical target \(\covartar^{(1)}\) or the analytical target~\(\covartar^{(2)}\) defined in section~\ref{subsec:inverted_linear_shrinkage_estimates} (hereafter referred to as simply the \textbf{covariance shrinkage estimate});
    \item the inverted \textbf{NERCOME estimate}, \(\estm{\prectrue}_\mathrm{NLS}\), introduced in section~\ref{subsec:non-linear_shrinkage} and constructed in section~\ref{subsec:non-linear_shrinkage_estimates};
    \item the direct precision linear-shrinkage estimate, \(\precsamp_{\mkskip\mathrm{LS}}\), introduced in section~\ref{subsec:direct_linear_shrinkage} with the empirical target \(\prectar^{(1)}\) or the analytical target~\(\prectar^{(2)}\) defined in section~\ref{subsec:direct_linear_shrinkage_estimates} (hereafter referred to as simply the \textbf{precision shrinkage estimate}).
\end{enumerate}

To this end, two types of performance metrics are considered: the first is based on the eigenvalue spectrum and loss function of the matrices themselves; the second is based on parameter inference, which is the ultimate objective of precision matrix estimation in cosmological likelihood analysis.
The benchmark used in all comparisons is the `reference' precision matrix~\(\precref\), which we recall from section~\ref{sec:application} is the sample estimate computed using all \num{2048} mock measurements and treated as a proxy for the true precision matrix~\(\prectrue\).

\subsection{Eigenvalue spectrum and loss function}

In general, since covariance and precision matrices are inversely related, smaller eigenvalues of the precision matrix correspond to larger inferred uncertainties, and the lowest eigenvalues correspond to the highest-variance principal components (eigenvectors) of the measurements.
The eigenvalue spectra of the different precision matrix estimates computed from subsets of \(\nsamp = \num{24}\) mock measurements are shown in Fig.~\ref{fig:eigenspectra_24}, and those from subsets of \(\nsamp = \num{30}\) mock measurements are shown in Fig.~\ref{fig:eigenspectra_30}.
For \(\nsamp = \num{2048}\) where all mock measurements are used in constructing the estimates (i.e. computed from the reference covariance matrix~\(\covarref\) wherever applicable), we show in Fig.~\ref{fig:eigenspectra_2048} the relative differences between the eigenvalue spectra instead.
We stress here that eigenvalues of different matrix estimates do not correspond to the same eigenvectors, and the ordering of the eigenvalues has been fixed to be monotonically increasing.
None the less, the eigenvalue spectrum can still serve as a useful diagnostic tool for assessing the performance of different estimators.
\begin{figure*}
    \settoheight{\subfigheight}{%
        \includegraphics[width=0.75\textwidth]{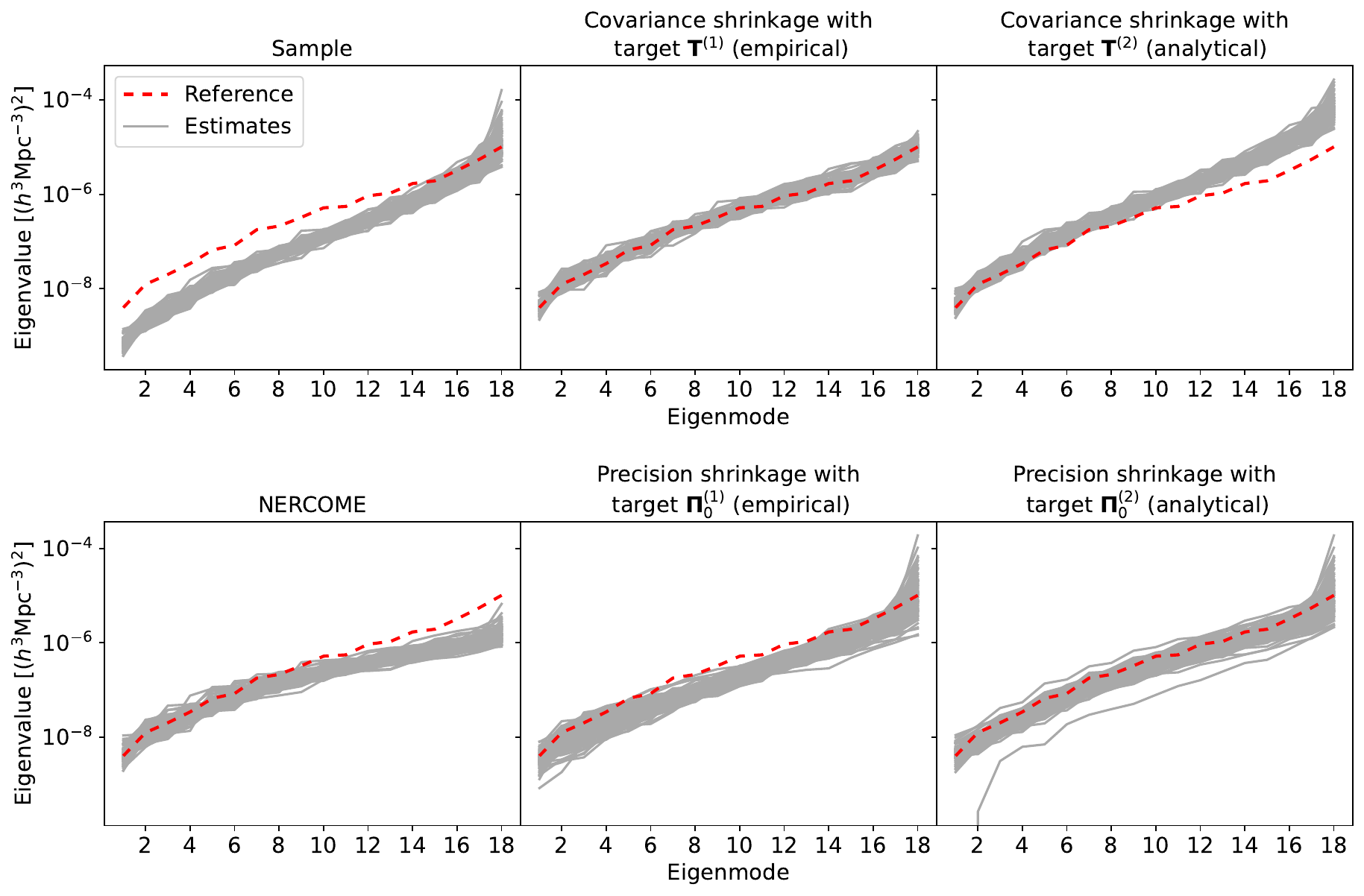}%
    }
    \includegraphics[width=0.75\textwidth]{graphics/figures/eigenspectra_n24}
    \caption{
        Eigenvalue spectra of the precision matrix estimates computed from subsets of \(\nsamp = \num{24}\) mock measurements.
        The eigenmodes are sorted in ascending order of eigenvalue.
        Each sub-figure shows the scatter of \num{85} spectra of one type of precision matrix estimate (with the target choice indicated where applicable), which are compared with the spectrum of the reference precision matrix~\(\precref\) shown by the dashed line.
    }
    \label{fig:eigenspectra_24}
\end{figure*}
\begin{figure*}
    \settoheight{\subfigheight}{%
        \includegraphics[width=0.75\textwidth]{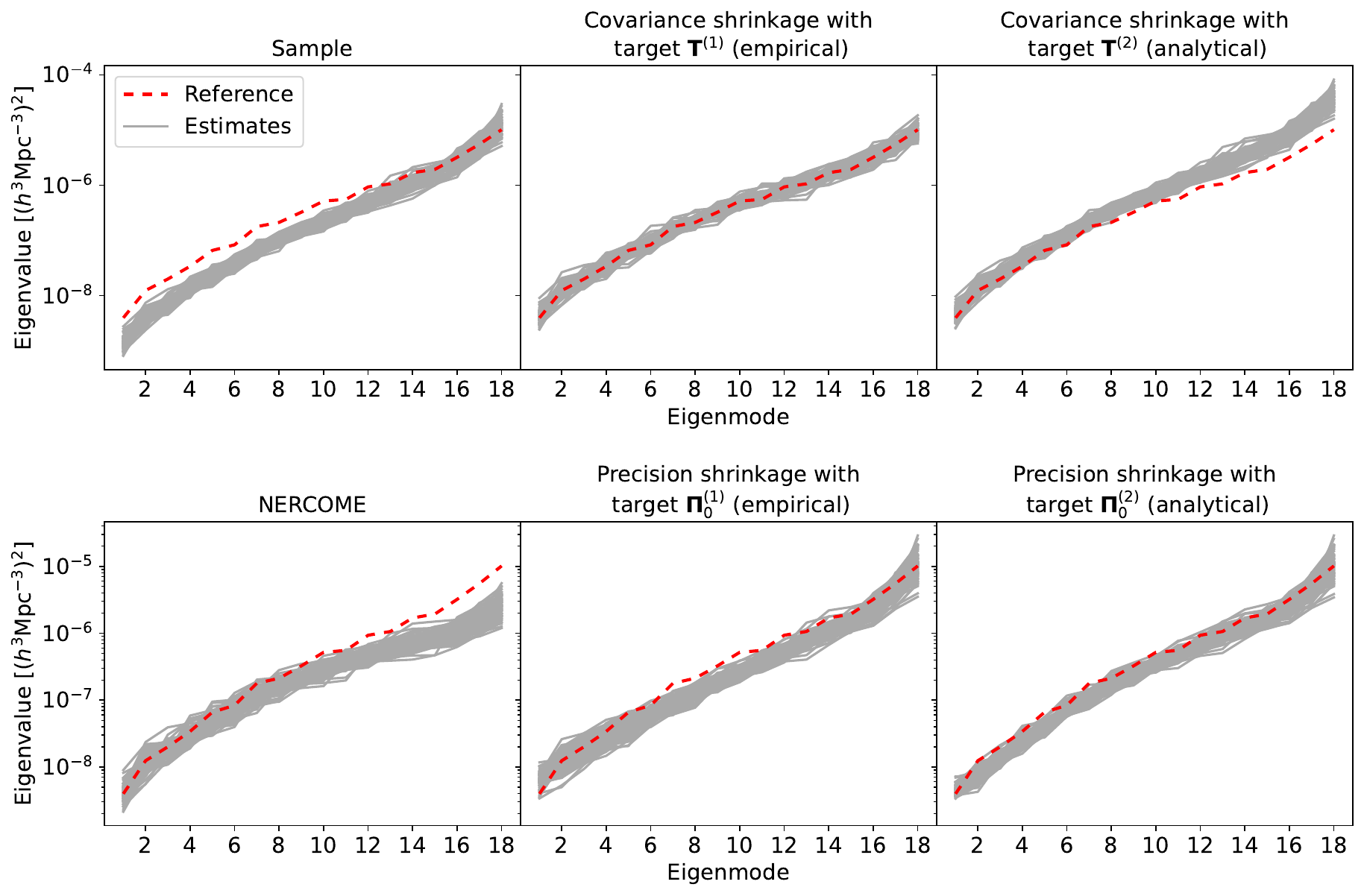}%
    }
    \includegraphics[width=0.75\textwidth]{graphics/figures/eigenspectra_n30}
    \caption{
        The same as Fig.~\ref{fig:eigenspectra_24} with eigenmodes sorted in ascending order of eigenvalue, but with \num{68} estimates computed from subsets of \(\nsamp = \num{30}\) mock measurements in each sub-figure.
    }
    \label{fig:eigenspectra_30}
\end{figure*}
\begin{figure*}
    \includegraphics[width=0.8\textwidth]{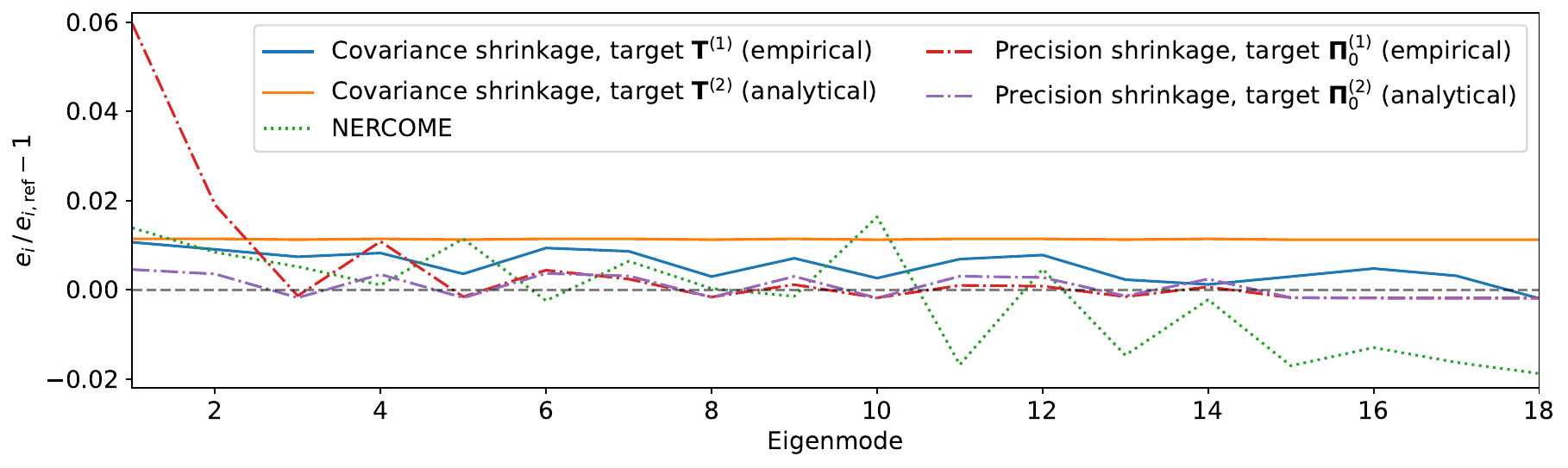}
    \caption{
        Relative difference, \(\ifrac{e_i}{e_{i,\mathrm{ref}}} - 1\), between the eigenvalue spectrum \(\{e_i\}_{i=1}^{\nddim}\) of each type of precision matrix estimate and that of the reference precision matrix, \(\{e_{i,\mathrm{ref}}\}_{i=1}^{\nddim}\), all computed using all \(\nsamp = \num{2048}\) mock measurements.
        The eigenmodes are sorted in ascending order of eigenvalue.
        The dashed grey line indicates the zero reference.
        Note that the inverted covariance shrinkage estimate~\(\estm{\prectrue}_\mathrm{LS}\) with target~\(\covartar^{(2)}\) (\textit{solid orange line}) is at a constant offset around \SI{1}{\percent}, as it is very close to the sample estimate but without the Hartlap factor, which is about \SI{1}{\percent} below unity.
    }
    \label{fig:eigenspectra_2048}
\end{figure*}

Overall, the eigenvalue spectra of the estimates approach that of the reference precision matrix~\(\precref\) as the sample size~\(\nsamp\) increases, with close alignment when \(\nsamp = \num{2048}\) as all estimates become close to the reference precision matrix itself.
There are several other observations of interest:
\begin{itemize}
    \item The sample estimates~\(\precsamp\) tend to lower the smallest eigenvalues, potentially leading to overestimated uncertainties;
    \item The covariance shrinkage estimates, \(\estm{\prectrue}_\mathrm{LS}\), with the target choice~\(\covartar^{(2)}\) tend to enlarge the largest eigenvalues, potentially leading to underestimated uncertainties. However, this does not occur for the target choice~\(\covartar^{(1)}\);
    \item The NERCOME estimates, \(\estm{\prectrue}_\mathrm{NLS}\), tend to decrease the largest eigenvalues, potentially leading to overestimated uncertainties;
    \item The precision shrinkage estimates~\(\precsamp_{\mkskip\mathrm{LS}}\) tend to slightly decrease the eigenvalues over the entire spectrum with the exception of the largest eigenvalues, which are more scattered.
    This could potentially lead to overestimated uncertainties.
\end{itemize}
It is also observed that when we use \(\prectar^{(2)}\) as target, one out of the \num{85} precision shrinkage estimates with \(\nsamp = \num{24}\) has negative eigenvalues, indicated by the grey line dropping towards zero in the bottom right plot in Fig.~\ref{fig:eigenspectra_24}.
We have noted that this precision shrinkage estimate with negative eigenvalues is also the only precision shrinkage estimate with a negative \(\opt{\estm{\alpha}}\)~value.
As explained in section~\ref{subsec:direct_linear_shrinkage}, negative \(\opt{\estm{\alpha}}\)~values can occur when the target~\(\prectar\) is too close to \(\covarsamp^{-1}\).
Since covariance/precision matrices should be positive semi-definite, we suggest substituting the target matrix with another one in such cases.
However, hereafter in this section we simply remove this spurious case, thus leaving 84 precision shrinkage matrix estimates~\(\precsamp_{\mkskip\mathrm{LS}}\) with analytical target~\(\prectar^{(2)}\).

We also note that for \(\nsamp = \num{2048}\), the eigenvalues of the covariance shrinkage estimate~\(\estm{\prectrue}_\mathrm{LS}\) with target~\(\covartar^{(2)}\) are at a constant offset around \SI{1}{\percent} from the reference eigenvalues, as indicated by the solid orange line in Fig.~\ref{fig:eigenspectra_2048}; this is because for this particular estimate, the shrinkage intensity parameter \(\opt{\estm{\lambda}} = \num{0.002}\) is close to zero, so that it is effectively the sample estimate itself albeit without the Hartlap factor that is about \SI{1}{\percent} below unity.

To make a more direct quantitative comparison, we consider the following  loss function for a generic precision matrix estimate~\(\estm{\prectrue}\):
\begin{equation}
    \label{eq:prec_loss_func}
    \func{L}\big(\estm{\prectrue}; \precref) = \fnorm*{\estm{\prectrue}^{1/2} \precref^{-1} \estm{\prectrue}^{1/2} - \mat{I}_\nddim} \,,
\end{equation}
where \(\mat{I}_\nddim\) is the \(\nddim\)-dimensional identity matrix.
We could have simply chosen the Frobenius distance~\(\fnorm{\estm{\prectrue} - \precref}\), but this loss function has the benefit of being physically dimensionless and thus slightly easier to interpret.
Over all subsets of \(\nsamp\) mock measurements where \(\nsamp = \num{24}\) or \num{30}, we compute the median loss function and its uncertainty bounds given by the \SI{16}{\percent} and \SI{84}{\percent} quantiles in Table~\ref{tab:precision_loss}, with additionally a single loss function value for \(\nsamp = \num{2048}\) as there is only one precision matrix estimate of each type using all available mock measurements.
\begin{table*}
    \caption{
        Median of the loss function~\(\func{L}\big(\estm{\prectrue})\) (equation~\ref{eq:prec_loss_func}) for each type of precision matrix estimate~\(\estm{\prectrue}\) with different sample sizes~\(\nsamp\), with uncertainty bounds given by the \SI{16}{\percent} and \SI{84}{\percent} quantiles.
        Note that there is only a single loss function value when \(\nsamp = \num{2048}\), as there is only one estimate of each type computed from all available mock measurements.
    }
    \begin{tabular*}{\textwidth}{%
        l@{\extracolsep{\fill}}r@{\extracolsep{4em}}r@{\extracolsep{4em}}r%
    }
        \toprule
        {Precision estimation type} & \(\nsamp = \num{24}\) & \(\nsamp = \num{30}\) & \(\nsamp = \num{2048}\) \\
        \midrule
        Sample & \meas{6.8}{2.2}{7.4} & \meas{5.4}{1.2}{1.8} & -- \\[4pt]
        Covariance shrinkage, target \(\covartar^{(1)}\) (empirical) & \meas{2.4}{0.3}{1.0} & \meas{2.3}{0.3}{0.6} & \num{0.051} \\[4pt]
        Covariance shrinkage, target \(\covartar^{(2)}\) (analytical) & \meas{38.3}{10.5}{25.2} & \meas{18.7}{4.3}{5.3} & \num{0.048} \\[4pt]
        NERCOME & \meas{3.6}{0.5}{1.2} & \meas{3.3}{0.4}{0.5} & \num{0.049} \\[4pt]
        Precision shrinkage, target \(\prectar^{(1)}\) (empirical) & \meas{7.2}{2.3}{7.8} & \meas{5.8}{1.6}{2.2} & \num{0.064} \\[4pt]
        Precision shrinkage, target \(\prectar^{(2)}\) (analytical) & \meas{6.6}{2.9}{8.7} & \meas{4.7}{1.1}{1.7} & \num{0.013} \\
        \bottomrule
    \end{tabular*}
    \label{tab:precision_loss}
\end{table*}

When the sample size is very close to the data vector dimension~\(\nddim = \num{18}\), e.g. \(\nsamp = \num{24}\), the covariance shrinkage estimate~\(\estm{\prectrue}_\mathrm{LS}\) with the empirical target~\(\covartar^{(1)}\) gives the smallest loss closely followed by the NERCOME estimate~\(\estm{\prectrue}_\mathrm{NLS}\).
The precision shrinkage estimate~\(\precsamp_{\mkskip\mathrm{LS}}\) with the analytical target~\(\prectar^{(2)}\) has a loss function close to that of the sample estimate~\(\precsamp\) while the other estimates result in larger losses.
When \(\nsamp = \num{30}\) mock measurements are used, most shrinkage estimates have a lower loss function than the sample estimate, except the covariance shrinkage estimate~\(\estm{\prectrue}_\mathrm{LS}\) with the analytical target~\(\covartar^{(2)}\) and the precision shrinkage estimate~\(\precsamp_{\mkskip\mathrm{LS}}\) with the empirical target~\(\prectar^{(1)}\).
When \(\nsamp = \num{2048}\), all loss function values are small.
Overall, both the covariance shrinkage estimate~\(\estm{\prectrue}_\mathrm{LS}\) with the empirical target~\(\covartar^{(1)}\) and the NERCOME estimate~\(\estm{\prectrue}_\mathrm{NLS}\) give consistently small loss values irrespective of the sample size, whereas the shrinkage estimates with an analytical target have larger loss function values but improve rapidly with the sample size~\(\nsamp\).

\subsection{Parameter inference}
\label{subsec:parameter_inference}

Since a major objective of precision matrix estimation is to obtain accurate cosmological parameter constraints in a likelihood analysis, in this section we consider the log-likelihood function from equation~\eqref{eq:likelihood} (up to an additive constant) for each type of precision matrix estimate~\(\estm{\prectrue}\):
\begin{multline}
    \ln\Like\big(b, f; \estm{\prectrue}) \\
    = - \frac{1}{2} \trs{\left[\vec{P}_\mathrm{data} - \vec{P}_\mathrm{model}(b, f) \right]} \estm{\prectrue} \left[\vec{P}_\mathrm{data} - \vec{P}_\mathrm{model}(b, f)\right] \,,
\end{multline}
where \(\vec{P}_\mathrm{data}\) is the concatenated data vector of power spectrum monopole, quadrupole and hexadecapole of the BOSS DR12 NGC catalogue, and \(\vec{P}_\mathrm{model}\) is the corresponding model vector given by equation~\eqref{eq:kaiser_model} transformed by the wide-angle and survey window matrices (see the beginning of section~\ref{sec:application} for details).
Note that for simplicity, we only consider varying the linear bias and growth-rate parameters~\(b\) and \(f\) with all other parameters fixed to the Planck 2015 cosmology~\citep{Planck2015}.
To perform Bayesian parameter inference, we assume uniform priors~\(b \sim \unif(\num{0.5}, \num{3.5})\) and \(f \sim \unif(\num{0}, \num{2})\), and sample the posterior distribution using {preconditioned Monte Carlo}~(PMC) implemented by the \codename{pocomc} sampler \citep{Karamanis2022,Karamanis2022:pocoMC}.
PMC uses a normal{\is}ing flow~\citep{Papamakarios2021} to remap the target distribution under consideration and then samples the resulting preconditioned distribution with an adaptive sequential Monte Carlo scheme \citep{DelMoral2006}.
We have chosen \codename{pocomc} for its ease of use and performance, but could have equally used more traditional Markov chain Monte Carlo~(MCMC) samplers such as \codename{emcee} and \codename{zeus} \citep{ForemanMackey2013,Karamanis2021}.

For each type of precision matrix estimate~\(\estm{\prectrue}\) computed from \(\nsamp = \num{24}\) mock measurements, we have obtained \num{85} chains of \num{2000} posterior samples each, except for the precision shrinkage estimates~\(\precsamp_{\mkskip\mathrm{LS}}\) with analytical target~\(\prectar^{(2)}\) which have 84 chains due to the removal of the single precision matrix estimate with negative eigenvalues.
For precision matrix estimates from \(\nsamp = \num{30}\) mock measurements, \num{68} posterior sample chains of the same length have been generated; and for \(\nsamp = \num{2048}\) where all mock measurements are used for precision matrix estimation, a single chain of at least \num{2000} posterior samples has been generated for each estimate type.

We first compare the marginal{\is}ed posterior distributions~\(\mathcal{P}_\theta\) of parameters \(\theta = b, f\) by considering their Kullback--Leibler~(KL) divergence \citep{Kullback1951} with respect to the reference posterior distribution obtained with the reference precision matrix estimate~\(\precref\),
\begin{equation}
    \label{eq:KL_div}
    \func{D_{\mathrm{KL}}}(\mathcal{P} \, \Vert \, \mathcal{P}_\mathrm{ref}) = \int_{-\infty}^{\infty} \mathcal{P}(x) \ln{\frac{\mathcal{P}(x)}{\mathcal{P}_\mathrm{ref}(x)}} \, \mathrm{d}x,
\end{equation}
where we have also used \(\mathcal{P}\) to denote the posterior density functions.
From each chain of posterior samples associated with a precision matrix estimate~\(\estm{\prectrue}\), we estimate the probability density function~\(\mathcal{P}\) using kernel density estimation by employing Gaussian kernels of bandwidth \num{0.1} \citep{Silverman1986}, so that equation~\eqref{eq:KL_div} may be evaluated.
Over all chains inferred from \(\nsamp\) mock measurements where \(\nsamp = \num{24}\) or \num{30}, we compute the median KL divergence \(D_{\mathrm{KL}}\) and its uncertainty bounds given by the \SI{16}{\percent} and \SI{84}{\percent} for both \(b\) and \(f\), which is presented in Table~\ref{tab:KL_divergences}; in addition, there is a single KL divergence value when \(\nsamp = \num{2048}\) for the posterior distribution obtained using all available mock measurements for precision matrix estimation.
\begin{table*}
    \caption{
        Median of the KL divergence value from the reference case for the marginal{\is}ed posterior distributions of bias~\(b\) and growth rate~\(f\), for each type of precision matrix estimate~\(\estm{\prectrue}\) with different sample sizes~\(\nsamp\), with uncertainty bounds given by the \SI{16}{\percent} and \SI{84}{\percent} quantiles.
        Note that there is only a single KL divergence value when \(\nsamp = \num{2048}\), as there is only one posterior of each type computed from all available mock measurements.
    }
    \begin{tabular*}{\textwidth}{l@{\extracolsep{\fill}}r@{\extracolsep{2.5em}}r@{\extracolsep{2.5em}}r@{\extracolsep{\fill}}r@{\extracolsep{2.5em}}r@{\extracolsep{2.5em}}r}
        \toprule
        & \multicolumn{3}{c}{\(D_{\mathrm{KL}}\) for \(b\)}  & \multicolumn{3}{c}{\(D_{\mathrm{KL}}\) for \(f\)} \\[4pt]
        Precision estimation type & \(\nsamp = \num{24}\) & \(\nsamp = \num{30}\) & \(\nsamp = \num{2048}\) & \(\nsamp = \num{24}\) & \(\nsamp = \num{30}\) & \(\nsamp = \num{2048}\) \\
        \midrule
        Sample & \meas{1.21}{0.95}{6.21} & \meas{0.51}{0.40}{1.11} & -- & \meas{0.79}{0.60}{7.10} & \meas{0.36}{0.28}{1.36} & -- \\[4pt]
        Covariance shrinkage, target \(\covartar^{(1)}\) (empirical) & \meas{0.05}{0.02}{0.09} & \meas{0.05}{0.02}{0.06} & \num{0.014} & \meas{0.08}{0.03}{0.05} & \meas{0.07}{0.03}{0.05} & \num{0.008} \\[4pt]
        Covariance shrinkage, target \(\covartar^{(2)}\) (analytical) & \meas{0.61}{0.33}{1.31} & \meas{0.47}{0.27}{0.78} & \num{0.009} & \meas{0.76}{0.42}{1.14} & \meas{0.40}{0.20}{0.55} & \num{0.012} \\[4pt]
        NERCOME & \meas{1.34}{0.88}{2.57} & \meas{0.54}{0.41}{1.25} & \num{0.018} & \meas{0.26}{0.21}{0.64} & \meas{0.21}{0.16}{0.37} & \num{0.018} \\[4pt]
        Precision shrinkage, target \(\prectar^{(1)}\) (empirical) & \meas{0.79}{0.66}{5.47} & \meas{0.29}{0.20}{1.05} & \num{0.010} & \meas{0.14}{0.10}{0.35} & \meas{0.11}{0.07}{0.29} & \num{0.006} \\[4pt]
        Precision shrinkage, target \(\prectar^{(2)}\) (analytical) & \meas{0.57}{0.46}{2.17} & \meas{0.27}{0.18}{0.84} & \num{0.010} & \meas{0.21}{0.08}{0.17} & \meas{0.13}{0.07}{0.14} & \num{0.013} \\
        \bottomrule
    \end{tabular*}
    \label{tab:KL_divergences}
\end{table*}

For sample size \(\nsamp = \num{24}\) or \num{30}, we see that the posterior distributions~\(\mathcal{P}_b\) for the bias parameter obtained from the sample estimates~\(\precsamp\) and the NERCOME estimates~\(\estm{\prectrue}_\mathrm{NLS}\) are most divergent from the reference.
For the growth-rate posterior distributions~\(\mathcal{P}_f\), the sample estimates~\(\precsamp\) and the covariance shrinkage estimates~\(\estm{\prectrue}_\mathrm{LS}\) with analytical target~\(\covartar^{(2)}\) lead to the greatest divergence from the reference.
The precision shrinkage estimates~\(\precsamp_{\mkskip\mathrm{LS}}\) with either target choice~\(\prectar^{(1)}\) or \(\prectar^{(2)}\) give posterior distributions that diverge less from the reference distribution than those obtained from the sample estimates~\(\precsamp\) when \(\nsamp = \num{24}\) or \num{30}.
For both sample sizes \(\nsamp = \num{24}\) and \num{30}, we obtain the smallest KL divergence values when we use the covariance shrinkage estimator~\(\estm{\prectrue}_\mathrm{LS}\) with empirical target~\(\covartar^{(1)}\) to infer posterior probability distributions.
When \(\nsamp = \num{2048}\), all divergence values are small indicating that the posterior distributions much more closely resemble the reference distribution.

Furthermore, in Figs.~\ref{fig:constraints_24} and \ref{fig:constraints_30}, we show the posterior contours of the \SI{68}{\percent} and \SI{95}{\percent} credible regions for a subset of three randomly chosen estimates of each precision matrix estimator type for sample size \(\nsamp = \num{24}\) and \num{30}.
In Fig.~\ref{fig:constraints_2048}, we show the single-posterior contours for each type of precision matrix estimate when \(\nsamp = \num{2048}\).
The posterior contours in those figures are compared to the reference results obtained using the reference precision matrix~\(\precref\), which yield the estimates \(b = \num{1.945+-0.018}\) and \(f = \num{0.617+-0.043}\).
The purpose of Figs.~\ref{fig:constraints_24}, \ref{fig:constraints_30} and \ref{fig:constraints_2048} is to offer a small subset of examples of inferred parameter constraints obtained using different precision matrix estimators for intuition and visual{\is}ation.
\begin{figure*}
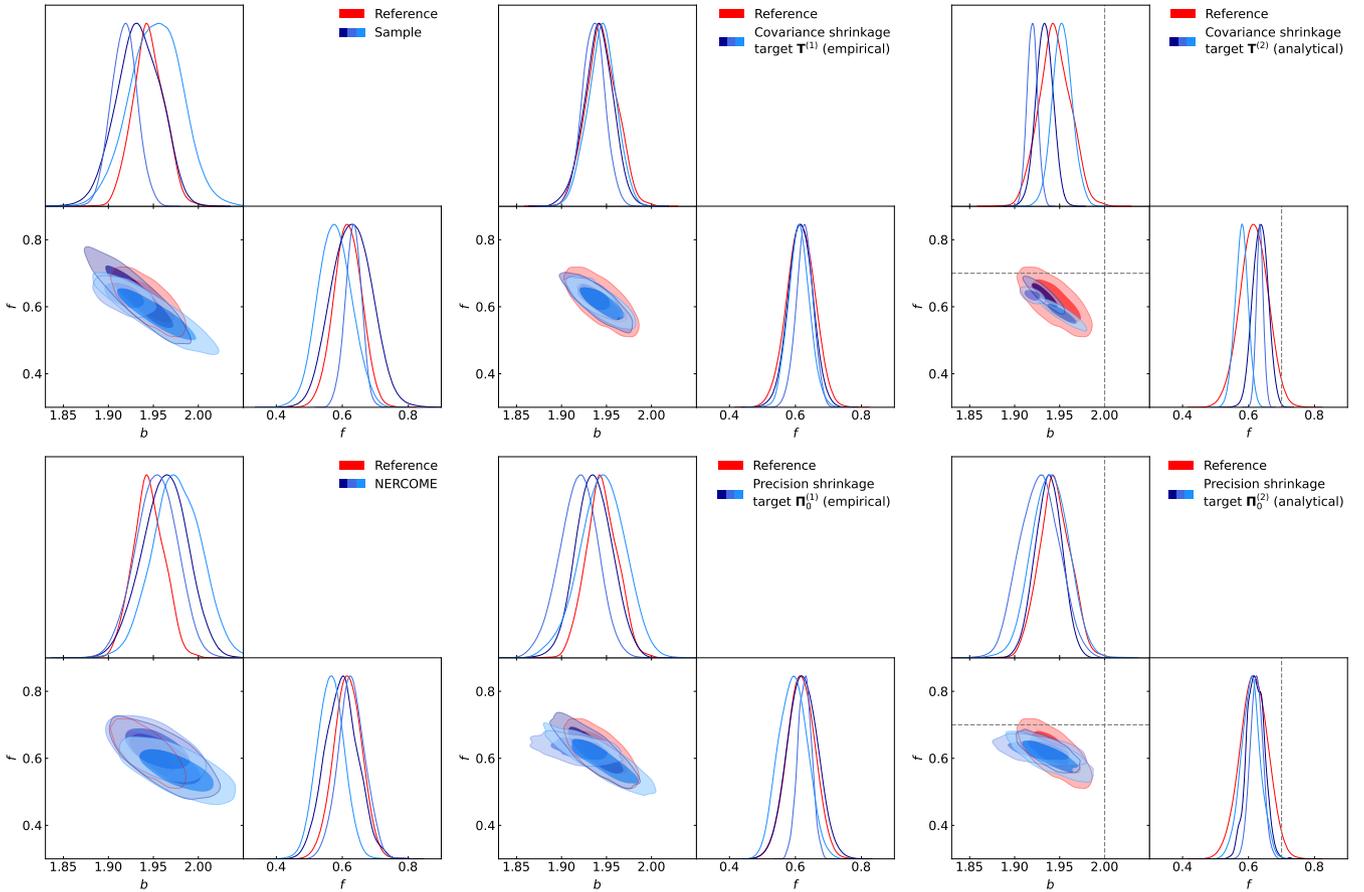

    \incgraphthree{figures/contours/cov_sample_n24}
    \incgraphthree{figures/contours/cov_shrinkage_emp_n24}
    \incgraphthree{figures/contours/cov_shrinkage_ana_n24}
    \incgraphthree{figures/contours/cov_NERCOME_n24}
    \incgraphthree{figures/contours/pre_shrinkage_emp_n24}
    \incgraphthree{figures/contours/pre_shrinkage_ana_n24}
    \caption{
        Posterior constraints in \SI{68}{\percent} and \SI{95}{\percent} credible regions on the linear bias~\(b\) and growth rate~\(f\) obtained from subsets of three randomly chosen precision matrix estimates of each type computed using \(\nsamp = \num{24}\) mock measurements (blue contours with three different hues), compared to the results obtained from the reference precision matrix~\(\precref\) (red contours).
        The dashed grey lines indicate the fiducial values \(b = \num{2}\) and \(f = \num{0.7}\) used to compute the analytical targets.
    }
    \label{fig:constraints_24}
\end{figure*}
\begin{figure*}
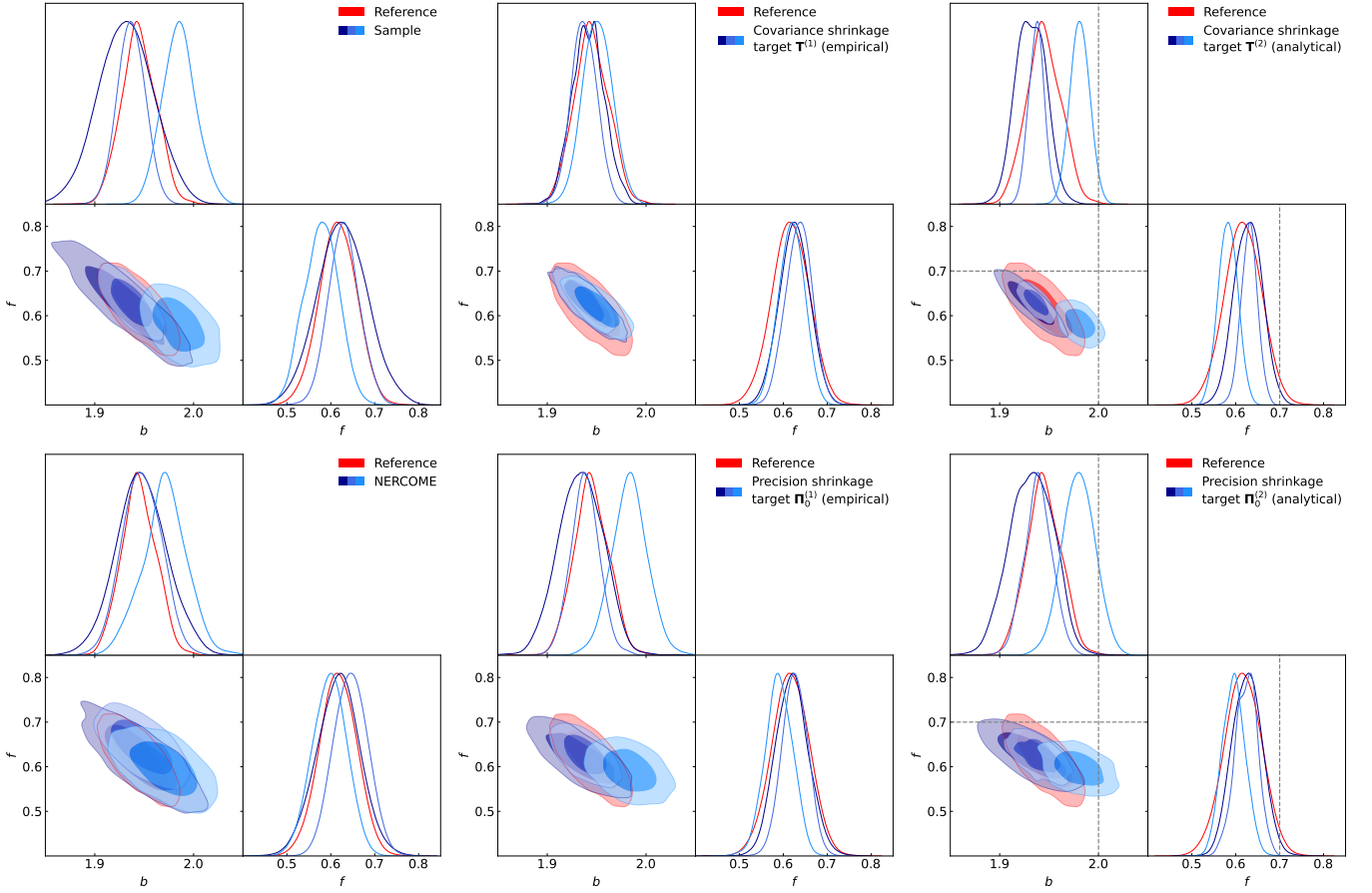

    \incgraphthree{figures/contours/cov_sample_n30}
    \incgraphthree{figures/contours/cov_shrinkage_emp_n30}
    \incgraphthree{figures/contours/cov_shrinkage_ana_n30}
    \incgraphthree{figures/contours/cov_NERCOME_n30}
    \incgraphthree{figures/contours/pre_shrinkage_emp_n30}
    \incgraphthree{figures/contours/pre_shrinkage_ana_n30}
    \caption{
        The same as Fig.~\ref{fig:constraints_24} but with precision matrix estimates computed from subsets of \(\nsamp = \num{30}\) mock measurements using the different estimators discussed in the text.
    }
    \label{fig:constraints_30}
\end{figure*}
\begin{figure*}
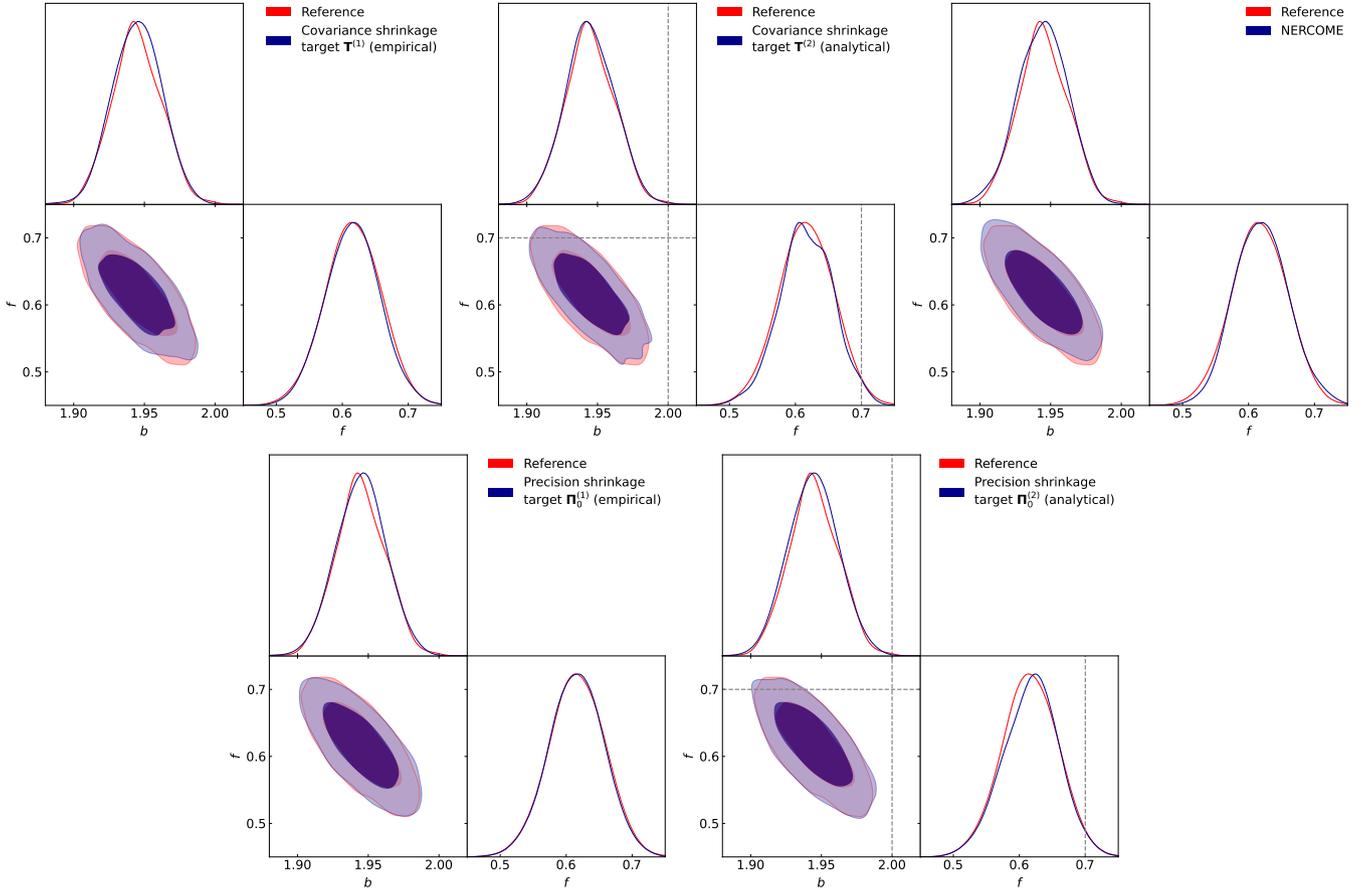

    \incgraphthree{figures/contours/cov_shrinkage_emp_n2048}
    \incgraphthree{figures/contours/cov_shrinkage_ana_n2048}
    \incgraphthree{figures/contours/cov_NERCOME_n2048}
    \incgraphthree{figures/contours/pre_shrinkage_emp_n2048}
    \incgraphthree{figures/contours/pre_shrinkage_ana_n2048}
    \caption{
        The same as Figs.~\ref{fig:constraints_24} and \ref{fig:constraints_30} but with precision matrix estimates computed from all \(\nsamp = \num{2048}\) mock measurements using the different shrinkage estimators discussed in the text.
    }
    \label{fig:constraints_2048}
\end{figure*}
In principle, the uncertainty in the precision matrix estimate will increase the uncertainties of the inferred parameters; for the sample precision matrix estimate, it is possible to derive a multiplicative factor that accounts for this in the uncertainties of the maximum likelihood estimator \citep{Dodelson2013}, which remains a good approximation for the posterior mean estimator when the posterior distribution is close to being Gaussian \citep{Percival2014}.
However, it is difficult to derive a similar factor for a generic shrinkage estimator as its probability distribution may not have an analytical form.

When the sample size is \(\nsamp = \num{24}\), which is small and close to the data vector dimension~\(\nddim = \num{18}\), we see in Fig.~\ref{fig:constraints_24} that the noisy sample estimator~\(\precsamp\) noticeably worsens the parameter constraints.
The covariance shrinkage estimator~\(\estm{\prectrue}_\mathrm{LS}\) with the empirical target~\(\covartar^{(1)}\) provides constraints much closer to the reference results, albeit very slightly overtightened, whereas the analytical target~\(\covartar^{(2)}\) leads to parameter uncertainties that are significantly underestimated.
The NERCOME estimator~\(\estm{\prectrue}_\mathrm{NLS}\) gives looser parameter constraints.
The precision shrinkage estimators~\(\precsamp_{\mkskip\mathrm{LS}}\) lead to broader contours that can have a very different orientation to the rest, and this gives overestimated uncertainties on \(b\), as well as slightly underestimated uncertainties on \(f\) especially for the analytical target~\(\prectar^{(2)}\).

With sample size \(\nsamp = \num{30}\), improvements are seen across all types of precision matrix estimators.
Again the covariance shrinkage estimator~\(\estm{\prectrue}_\mathrm{LS}\) with the empirical target~\(\covartar^{(1)}\) provides posteriors closest to the reference results, albeit with slightly over-tightened parameter constraints as before.
The covariance shrinkage estimator~\(\estm{\prectrue}_\mathrm{LS}\) with the analytical target~\(\covartar^{(2)}\) still leads to underestimated parameter constraints, but the effect is less severe than for sample size \(\nsamp = \num{24}\).
As for the precision shrinkage estimators~\(\precsamp_{\mkskip\mathrm{LS}}\), the orientation of the posterior contours is less rotated from the rest than for \(\nsamp = \num{24}\).

When computed from all \num{2048} mock measurements, all types of precision matrix estimates yield posterior constraints in close approximation to those obtained from using the reference precision matrix~\(\precref\), as seen in Fig.~\ref{fig:constraints_2048}.
This indicates that given a large sample size, all estimation methods behave consistently as expected.

To quantify the differences in the posterior constraints with a varying sample size~\(\nsamp\) (as shown in Figs.~\ref{fig:constraints_24}, \ref{fig:constraints_30} and \ref{fig:constraints_2048}), we compute the mean \(\estm{b}\) or \(\estm{f}\) and the corresponding standard deviation~\(\estm{\sigma}\) of the marginal posterior distributions from the combined chain for each type of precision matrix estimate; these are then compared against the reference posterior mean \(\est{b}_\mathrm{ref}\) or \(\est{f}_\mathrm{ref}\) and standard deviation~\(\est{\sigma}_\mathrm{ref}\) extracted from the chain obtained with the reference precision matrix, as shown in Fig.~\ref{fig:inferred_params}.
\begin{figure*}
    \includegraphics[width=0.95\textwidth]{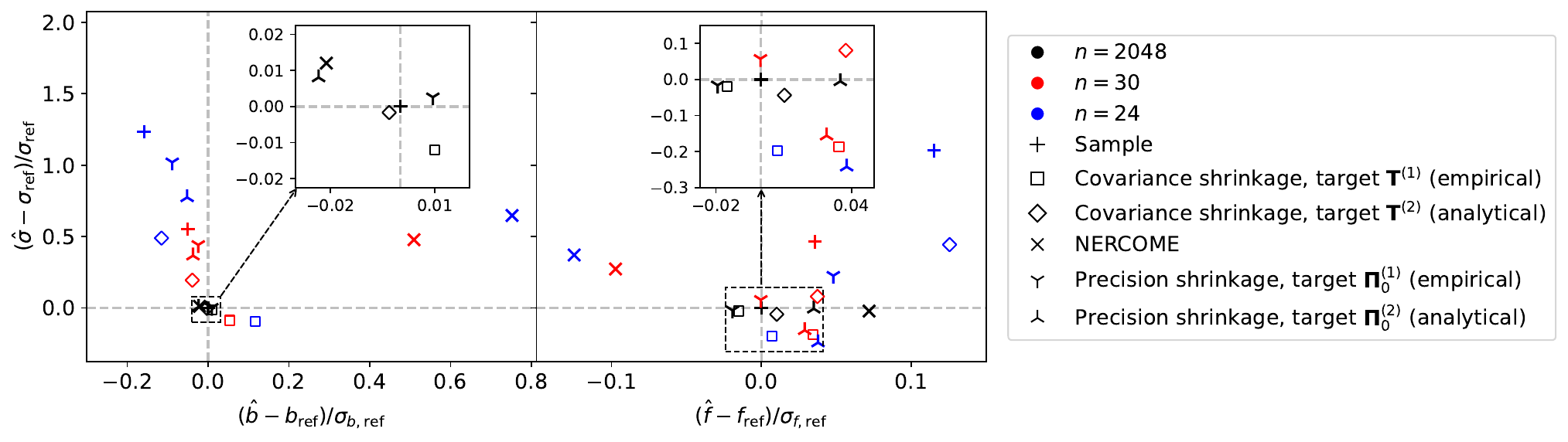}
    \caption{
        Relative deviations in the posterior mean and standard deviation of the linear bias~\(b\) (\textit{left panel}) and growth rate~\(f\) (\textit{right panel}) obtained from the various precision matrix estimates with a varying sample size~\(\nsamp\), compared to the reference results~\(b_\mathrm{ref}\) or \(f_\mathrm{ref}\) and \(\sigma_\mathrm{ref}\) obtained using the reference precision matrix.
        The dashed grey lines indicate the zero reference.
    }
    \label{fig:inferred_params}
\end{figure*}
It is visually clear from Fig.~\ref{fig:inferred_params} that when \(\nsamp\) is small, the NERCOME estimate leads to a noticeable shift in the posterior mean for the linear bias~\(b\) and the sample estimate leads to the most overestimated parameter uncertainties~\(\estm{\sigma}\), while both the covariance shrinkage and precision shrinkage estimates generally produce posterior means that are closer to the reference.
Furthermore, the covariance shrinkage estimates also lead to parameter uncertainties~\(\estm{\sigma}\) that are significantly closer to the reference~\(\est{\sigma}_\mathrm{ref}\) than those obtained from the sample estimates for \(\nsamp=\num{24}\) or \num{30}.
In contrast, the precision shrinkage estimates result in parameter uncertainties that are only marginally better than those obtained from the sample estimates, especially in the case of linear bias~\(b\).
This is perhaps related to the observation that precision shrinkage tends to underestimate the eigenvalues of the precision matrix compared to covariance shrinkage, though to a lesser extent than sample estimation (see Figs.~\ref{fig:eigenspectra_24} and \ref{fig:eigenspectra_30}).
Finally, we notice that when the sample size is \(\nsamp = \num{2048}\), all estimates produce results that are close to the reference case, with the exception of the NERCOME estimate for the growth rate~\(f\), where we again observe a shift in the posterior mean.

\Section{Conclusion}
\label{sec:conclusion}

The determination of the covariance and precision matrices is vital to parameter inference in the era of precision cosmology.
In this work, we have focused on a class of estimation methods known as shrinkage, which are used to obtain the following precision matrix estimates:
\begin{enumerate}
    \item the covariance shrinkage estimate~\(\estm{\prectrue}_\mathrm{LS}\) obtained after inversion with two target choices~\(\covartar^{(1)}\) and \(\covartar^{(2)}\), where the former is empirically based on the diagonal of the sample covariance matrix estimate and the latter is analytical (see section~\ref{subsec:inverted_linear_shrinkage_estimates});
    \item the NERCOME (non-linear shrinkage) estimate~\(\estm{\prectrue}_\mathrm{NLS}\) obtained after inversion (see section~\ref{subsec:non-linear_shrinkage_estimates});
    \item the precision shrinkage estimate~\(\est{\precsamp}_{\mkskip\mathrm{LS}}\) with two target choices~\(\prectar^{(1)}\) and \(\prectar^{(2)}\), where the former is empirically based on the eigenvalues of the inverted sample covariance matrix estimate and the latter is analytical (see section~\ref{subsec:direct_linear_shrinkage_estimates}).
\end{enumerate}
All of these shrinkage estimates, as well as the sample estimate~\(\est{\precsamp}\), are computed for a varying sample size~\(\nsamp\) and compared to the reference precision matrix~\(\precref\) obtained for a large value of \(\nsamp\).

We have applied these estimation methods to the power spectrum analysis of the BOSS DR12 data set with data vector dimension~\(\nddim = \num{18}\), in conjunction with \num{2048} \codename{patchy} mock catalogues for precision matrix estimation.
The performance of the different precision matrix estimates for sample size~\(\nsamp = \num{24}\), \num{30} or \num{2048} is compared against the reference precision matrix~\(\precref\) constructed using all mock measurements.
We have considered diagnostic metrics such as the matrix loss function and the eigenvalue spectrum; in addition, we have also performed Bayesian parameter inference of the linear bias and growth-rate parameters~\(b\) and \(f\), and calculated the KL divergence of the inferred marginal posterior distributions with respect to the reference distributions obtained using \(\precref\). It has been found that:
\begin{itemize}
    \item When the sample size is small, the covariance shrinkage estimate~\(\estm{\prectrue}_\mathrm{LS}\) with the empirical target~\(\covartar^{(1)}\) provides the lowest loss function values, the closest eigenvalue spectrum, the smallest KL divergence values and the best agreement in the posterior parameter constraints compared to the reference results. With the analytical target~\(\covartar^{(2)}\), however, it has a higher loss function value, higher KL divergence values and less consistent parameter constraints.
    \item The NERCOME estimate~\(\estm{\prectrue}_\mathrm{NLS}\) tends to underestimate the largest precision matrix eigenvalues leading to weaker parameter constraints, which has also been demonstrated by \citet{GouyouBeauchamps2023}. It also produces KL divergence values which are as large as those produced by the sample estimate~\(\precsamp\) for the bias parameter~\(b\).
    \item The precision shrinkage estimate~\(\est{\precsamp}_{\mkskip\mathrm{LS}}\) produces parameter constraints with a slightly rotated parameter degeneracy direction for small sample size~\(\nsamp = \num{24}\) or \num{30}, but still has lower KL divergence values than the sample estimate~\(\precsamp\).
    \item As expected, when the sample size~\(\nsamp\) is very large, all precision estimates behave similarly and give consistent posterior constraints.
\end{itemize}
Based on these configurations and results, we suggest the use of covariance shrinkage estimation of the precision matrix obtained after inversion.
This serves as a useful guide for the cosmological analyses of forthcoming data releases of Stage-IV galaxy surveys such as DESI and \textit{Euclid}, where large data sets are best exploited with more performant precision matrix estimates.
However, it is worth expanding upon this proof-of-concept study by including more non-linear clustering scales and considering non-Gaussian summary statistics and covariance matrices.
We leave these interesting extensions to future work.

Finally, we note there are many other possible choices of the shrinkage target not exhausted in this work, and their performance will depend on the precise configuration of individual cosmological analyses.
In particular, although the precision shrinkage estimate has been found to give potentially worse parameter constraints with a rotated parameter degeneracy direction, it needs not be the case with other unexplored target choices or in a different analysis set-up.
It is expected that with a target matrix that is more structurally similar to the true precision matrix \(\prectrue\) (instead of the diagonal targets considered in section~\ref{subsec:direct_linear_shrinkage_estimates}), e.g. an analytical target with non-linear and window function effects, these precision shrinkage estimates can be further improved.
Similarly, it would be interesting to explore intermediate sample sizes for covariance matrix estimation; for instance, moderately larger values \(n = \text{\numrange[range-phrase=--]{50}{100}}\) for the data vector dimension~\(\nddim = 18\) in this work, perhaps with a greater number of total mock catalogues than the available \num{2048}.
Moreover, it would offer new insight to derive correction factors similar to those in \citet{Hartlap2006} and \citet{Percival2014} to account for the stochasticity in shrinkage estimates of the precision matrix, even when they are already much less noisy than the standard sample estimates.
However, this is generally difficult since shrinkage estimates are composite quantities for which the probability distribution may be analytically intractable.
We also leave these considerations to future work.

{
\Section*{Acknowledgements}

The authors would like to thank Minas Karamanis for inspiration; Uro\v{s} Seljak, Richard Grummit, Benjamin Joachimi, Julian Bautista and Taras Bodnar for helpful discussions; and the anonymous referee for providing constructive feedback that has improved the manuscript in many aspects.

This project has received funding from the European Research Council (ERC) under the European Union’s Horizon 2020 research and innovation programme (grant agreement 853291). MJL acknowledges support from the School of Physics \& Astronomy at the University of Edinburgh. FB acknowledges the support of the Royal Society through a University Research Fellowship.

This work has made use of publicly available \codename{python} packages
\codename{numpy} \citep{Harris2020},
\codename{scipy} \citep{SciPy2020},
\codename{matplotlib} \citep{Hunter2007},
\codename{getdist} \citep{Lewis2019},
\codename{nbodykit} \citep{Hand2018}
and
\codename{pocomc} \citep{Karamanis2022:pocoMC}.

}

{
\hypertarget{data_availability}{}%
\Section*{Data Availability}

The data underlying this article will be shared on request to the corresponding author.
The source data are available at \weblink{https://fbeutler.github.io/hub/deconv_paper}{fbeutler.github.io/hub/deconv\_paper}, and any generated data, including the code used to derive further data, are available at \weblink{https://github.com/marnixlooijmans/shrinkage-estimation-paper}{github.com/marnixlooijmans/shrinkage-estimation-paper}.

}

{
\raggedright
\hypersetup{urlcolor=MNRASPurple}
\bibliographystyle{mnras}
\bibliography{references}
}

\bsp

\vfill

\makeatletter
\epeven@declaration
\vspace{-10pt}
\makeatother

\label{lastpage}
\end{document}

%% file: preamble.tex
\usepackage{amsmath}
\usepackage{booktabs}               
\usepackage{bm}                     
\usepackage[calc,en-GB]{datetime2}  
\usepackage{etoolbox}
\usepackage[T1]{fontenc}
\usepackage{graphicx}
\usepackage{mathtools}              
\usepackage{microtype}              
\usepackage{newtxtext}
\usepackage[slantedGreek,vvarbb]{newtxmath}
\usepackage{siunitx}                
\usepackage{xcolor}                 
\usepackage{xparse}                 

\usepackage[
    only,llbracket,rrbracket
]{stmaryrd}                         

\DeclareMathAlphabet{\mathsfbf}{OT1}{\sfdefault}{b}{n}
\SetMathAlphabet{\mathsfbf}{bold}{OT1}{\sfdefault}{b}{n}

\graphicspath{{graphics/}}

\newdimen\subfigheight

\definecolor{MNRASPurple}{HTML}{BC3B9C}

\DTMnewdatestyle{MNRASDate}{%
}
\DTMnewdatestyle{MNRASYear}{%
}
\DTMsetdatestyle{MNRASDate}

\ExplSyntaxOn
\DeclareExpandableDocumentCommand{\IfNoValueOrEmptyTF}{mmm}{
    \IfNoValueTF{#1}{#2}{{\tl_if_empty:nTF {#1} {#2} {#3}}}
}
\ExplSyntaxOff

\DeclareRobustCommand{\SURNAME}[3]{#2}
\let\originalbibliography\thebibliography
\renewcommand{\thebibliography}{%
    \DeclareRobustCommand{\SURNAME}[3]{##3}\originalbibliography%
}

\makeatletter
\patchcmd{\NAT@citex}{%
    \@citea\NAT@hyper@{%
        \NAT@nmfmt{\NAT@nm}%
        \hyper@natlinkbreak{\NAT@aysep\NAT@spacechar}{\@citeb\@extra@b@citeb}%
        \NAT@date%
    }%
}{%
    \@citea\NAT@nmfmt{\NAT@nm}\NAT@aysep\NAT@spacechar\NAT@hyper@{\NAT@date}%
}{}{}
\patchcmd{\NAT@citex}{%
    \@citea\NAT@hyper@{%
        \NAT@nmfmt{\NAT@nm}%
        \hyper@natlinkbreak{%
            \NAT@spacechar\NAT@@open\if*#1*\else#1\NAT@spacechar\fi%
        }{%
            \@citeb\@extra@b@citeb%
        }%
        \NAT@date%
    }%
}{%
    \@citea\NAT@nmfmt{\NAT@nm}%
    \NAT@spacechar\NAT@@open\if*#1*\else#1\NAT@spacechar\fi%
    \NAT@hyper@{\NAT@date}%
}{}{}
\makeatother

\makeatletter
\newcommand\footnoteref[1]{\protected@xdef\@thefnmark{\ref{#1}}\@footnotemark}
\makeatother

\newcommand{\affil}[2]{\hyperlink{#1}{\textsuperscript{#2}}}
\newcommand{\correspondence}[1]{%
    \thanks{\hspace{-0.5em} Email: \href{mailto:#1}{{#1}}}%
}
\newcommand{\orcid}[1]{%
    \hspace{0.1\baselineskip}%
    \raisebox{-0.1\baselineskip}{%
        \href{https://orcid.org/#1}{%
            \includegraphics[height=0.7\baselineskip,keepaspectratio]%
                {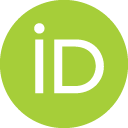}%
        }%
    }%
}

\NewDocumentCommand{\addauthor}{smmmoo}{%
    #2
    \IfNoValueOrEmptyTF{#5}{}{\,\orcid{#5}}
    \IfBooleanTF{#1}{}{,}
    \affil{#3}{#4}
    \IfNoValueTF{#6}{}{\correspondence{#6}}
    \IfBooleanTF{#1}{}{ }
}

\newcommand{\addaffil}[3]{\hypertarget{#1}{\textsuperscript{\textup{#2}}}#3}

\NewDocumentCommand{\Section}{sm}{%
    \IfBooleanTF{#1}{%
        \section*{\texorpdfstring{\textls{#2}}{#2}}%
    }{%
        \section{\texorpdfstring{\textls{#2}}{#2}}%
    }%
}

\newcommand{\incgraphthree}[1]{\includegraphics[width=0.33\textwidth]{#1}}

\newcommand{\arxivprint}[1]{%
    ({\hypersetup{urlcolor=blue}\eprint{arXiv}{#1}})%
}                                              
\newcommand{\weblink}[2]{\href{#1}{{#2}}}      
\newcommand{\codename}[1]{\textsc{#1}}         

\newcommand{\Jrn}{J.~}  
\newcommand{\is}{iz}    

\DeclareSIUnit{\parsec}{pc}               
\DeclareSIUnit{\h}{\text{$h$}}            
\DeclareSIUnit{\stdsig}{\text{$\sigma$}}  

\DeclarePairedDelimiter{\fnorm}{\Vert}{\Vert_\mathrm{F}}      
\DeclarePairedDelimiter{\tnorm}{\Vert}{\Vert_\mathrm{tr}}     

\makeatletter
\DeclareDocumentCommand{\@arguments}{
    t\big t\Big t\bigg t\Bigg g o d() d||
}{
    \IfBooleanTF{#1}{%
        \let\ltag\bigl \let\rtag\bigr
    }{%
        \IfBooleanTF{#2}{%
            \let\ltag\Bigl \let\rtag\Bigr
        }{%
            \IfBooleanTF{#3}{%
                \let\ltag\biggl \let\rtag\biggr
            }{%
                \IfBooleanTF{#4}{%
                    \let\ltag\Biggl \let\rtag\Biggr
                }{%
                    \let\ltag\left \let\rtag\right
                }%
            }%
        }%
    }%
    \IfNoValueTF{#5}{%
        \IfNoValueTF{#6}{%
            \IfNoValueTF{#7}{%
                \IfNoValueTF{#8}{%
                    ()
                }{%
                    \ltag\lvert{#8}\rtag\rvert
                }%
            }{%
                \ltag(#7\rtag) \IfNoValueTF{#8}{}{|#8|}
            }%
        }{%
            \ltag[#6\rtag]
            \IfNoValueTF{#7}{}{(#7)}
            \IfNoValueTF{#8}{}{|#8|}
        }%
    }{%
        \ltag\lbrace#5\rtag\rbrace
        \IfNoValueTF{#6}{}{[#6]}
        \IfNoValueTF{#7}{}{(#7)}
        \IfNoValueTF{#8}{}{|#8|}
    }%
    \ifnum\z@=`{\fi}
}
\makeatother

\DeclareDocumentCommand{\argopen}{s}{%
    \IfBooleanTF{#1}{%
        \mathopen{}\mathclose\bgroup%
    }{%
        \mathopen{}\mathclose\bgroup\left%
    }
}
\DeclareDocumentCommand{\argclose}{s}{%
    \IfBooleanTF{#1}{\egroup}{\aftergroup\egroup\right}%
}

\makeatletter
\let\originalleft\left
\renewcommand{\left}{\mathopen{}\mathclose\bgroup\originalleft}
\let\originalright\right
\renewcommand{\right}{\aftergroup\egroup\originalright}
\DeclareDocumentCommand{\oper}{s m}{%
    \IfBooleanTF{#1}{%
        \operatorname{#2}%
    }{%
        \operatorname{#2}\!{\ifnum\z@=`}\fi\@arguments%
    }%
}
\DeclareDocumentCommand{\func}{s m}{%
    \IfBooleanTF{#1}{%
        #2%
    }{%
        \mathop{}\!#2{\ifnum\z@=`}\fi\@arguments%
    }%
}
\makeatother

\newcommand{\mkskip}{\mkern2mu}  

\makeatletter
\newcommand{\transT}{{\mathpalette\@transT{}}}
\newcommand{\@transT}[2]{\raisebox{\depth}{$\m@th#1\intercal$}}
\makeatother  

\newcommand{\trs}[1]{{#1}^\transT}   
\newcommand{\mean}[1]{\widebar{#1}}  
\newcommand{\estm}[1]{\hat{#1}}      
\newcommand{\est}[1]{#1}             
\newcommand{\opt}[1]{#1^{*}}         

\newcommand{\real}{\mathbb{R}}            
\newcommand{\unif}{\mathcal{U}}           

\renewcommand{\vec}[1]{\mathbfit{#1}}  
\newcommand{\mat}[1]{\mathsfbf{#1}}    

\NewDocumentCommand{\ifrac}{o m m}{%
    \IfNoValueTF{#1}{{%
        \argopen.\mkern-2mu{#2}\mkern2mu%
        \middle/%
        \mkern2mu{#3}\mkern-2mu\argclose.%
    }}{#2\,#1|\,#3}%
}  
\NewDocumentCommand{\intrange}{o m m}{%
    \IfNoValueTF{#1}{%
        \ensuremath{\left\llbracket#2,#3\right\rrbracket}%
    }{\ensuremath{\llbracket#2,#3\rrbracket}}%
}  

\NewDocumentCommand{\tr}{s}{%
    \IfBooleanTF#1{\oper*{tr}}{\oper{tr}}%
}  
\NewDocumentCommand{\diag}{s}{%
    \IfBooleanTF#1{\oper*{diag}}{\oper{diag}}%
}  

\NewDocumentCommand{\Exp}{s}{%
    \IfBooleanTF#1{\oper*{\mathbb{E}}}{\oper{\mathbb{E}}}%
}  
\NewDocumentCommand{\Var}{s}{%
    \IfBooleanTF#1{\oper*{Var}}{\oper{Var}}%
}  
\NewDocumentCommand{\Cov}{s}{%
    \IfBooleanTF#1{\oper*{Cov}}{\oper{Cov}}%
}  
\NewDocumentCommand{\eVar}{s}{%
    \IfBooleanTF#1{%
        \oper*{\ensuremath{\widehat{\mathrm{Var}}}}%
    }{%
        \oper{\ensuremath{\widehat{\mathrm{Var}}}}%
    }%
}  
\NewDocumentCommand{\eCov}{s}{%
    \IfBooleanTF#1{%
        \oper*{\ensuremath{\widehat{\mathrm{Cov}}}}%
    }{%
        \oper{\ensuremath{\widehat{\mathrm{Cov}}}}%
    }%
}  

\NewDocumentCommand{\Like}{s o}{%
    \IfBooleanTF#1{
        \IfNoValueTF{#2}{%
            \func*{\mathcal{L}}%
        }{%
            \func*{\mathcal{L}_{#2}}%
        }%
    }{%
        \IfNoValueTF{#2}{%
            \func{\mathcal{L}}%
        }{%
            \func{\mathcal{L}_{#2}}%
        }%
    }%
}  

\newcommand{\meas}[3]{${#1}\substack{+{#3}\\-{#2}}$}  
                                                           
\newcommand{\nddim}{d}  
\newcommand{\nsamp}{n}  
\newcommand{\npara}{p}  

\newcommand{\datamat}{\mat{X}}               
\newcommand{\covartrue}{\mat{\Sigma}}        
\newcommand{\prectrue}{\mat{\Psi}}           
\newcommand{\covarsamp}{\est{\mat{S}}}       
\newcommand{\precsamp}{\est{\mat{\Pi}}}      
\newcommand{\covartar}{\mat{T}}              
\newcommand{\prectar}{\mat{\Pi}_{\mkskip0}}  
\newcommand{\covarref}{%
    \est{\mat{S}}_\mathrm{ref}%
}                                            
\newcommand{\precref}{%
    \est{\mat{\Pi}}_{\mkskip\mathrm{ref}}%
}                                            

\newcommand{\hfac}{\gamma}